\documentclass[showpacs,amsmath,amssymb]{revtex4}

\usepackage{graphicx}
\usepackage{dcolumn}
\usepackage{bm}

\begin{document}



\title{CRITICALITY ON  NETWORKS WITH TOPOLOGY-DEPENDENT INTERACTIONS}

\author{C.V. Giuraniuc$^{1}$, J.P.L. Hatchett$^{2}$, J.O. Indekeu$^{1}$, M.
Leone$^{3}$, I. P\'erez Castillo$^{4}$, B. Van Schaeybroeck$^{1}$,
C. Vanderzande$^{4,5}$}
\affiliation{%
$^{1}$Laboratorium voor Vaste-Stoffysica en Magnetisme, Katholieke
Universiteit Leuven, 3001 Leuven, Belgium,\\$^{2}$Laboratory for
Mathematical Neuroscience, RIKEN Brain Science Institute,Wako-Shi,
Saitama 351-0198, Japan
\\$^{3}$Institute for Scientific Interchange (I.S.I.),
Villa Gualino, Viale Settimio Severo 65, 10133 Turin, Italy,\\
$^{4}$Rudolf Peierls Center for Theoretical Physics,University of
Oxford, 1 Keble Road, Oxford, OX1 3NP, UK\\$^{5}$Departement WNI,
Hasselt University Agoralaan, Gebouw D, B-3590 Diepenbeek, Belgium
}%


\date{\today}

\begin{abstract}
Weighted scale-free networks with topology-dependent interactions
are studied. It is shown that the possible universality classes of
critical behaviour, which are known to depend on topology, can
also be explored by tuning the form of the interactions at fixed
topology. For a model of opinion formation, simple mean field and
scaling arguments show that a mapping
$\gamma'=(\gamma-\mu)/(1-\mu)$ describes how a shift of the
standard exponent $\gamma$ of the degree distribution can absorb
the effect of degree-dependent pair interactions $J_{ij} \propto
(k_ik_j)^{-\mu}$, where $k_i$ stands for the degree of vertex $i$.
This prediction is verified by extensive numerical investigations
using the cavity method and Monte Carlo simulations. The critical
temperature of the model is obtained through the Bethe-Peierls
approximation and with the replica technique. The mapping can be
extended to nonequilibrium models such as those describing the
spreading of a disease on a network.
\end{abstract}

\maketitle

\vspace*{1pt}
\section{Introduction}
\vspace*{-0.5pt}
\noindent
In recent years \cite{-2,-1} it has become clear that many natural and technological networks are quite different from simple random networks \cite{-5} and share several unexpected properties such as a scale-free degree distribution \cite{-2,-1}, small world connectivity \cite{-4}, soft modularity\footnote{By soft-modularity, one means that the network consists of different modules whose mutual interactions are suppressed but not completely eliminated \cite{-3}.}\   and so on. Much work has gone into a precise characterisation of these {\it topological} properties for a variety of networks as diverse as the internet \cite{101}, metabolic networks in cells \cite{102} or networks of chemical reactions in planetary atmospheres \cite{104}.
Nontrivial topology is now established as an essential ingredient of complex systems \cite{103}. This observation naturally raises the question how these structures are formed and grow. A well known mechanism is that of preferential attachment which leads to the Barab\'{a}si-Albert network with a power-law degree distribution \cite{1}.

Another important question is how the network topology affects
physical properties such as collective behaviour, transport
quantities, the spreading of a disturbance, ... Particular
interest has been devoted to the behaviour of the Ising model on a
network. Besides being the standard model of (equilibrium)
statistical mechanics, the Ising model is also expected to give a
simple description of sociological phenomena such as opinion
formation \cite{105}. The first studies of this model indicated
\cite{0} that on a Barab\'{a}si-Albert network, the Ising model is
always ordered. It was, however, soon realised that a finite
transition temperature can be obtained if one considers the Ising
model on scale-free networks with a degree distribution that is
less long ranged than that of the Barab\'{a}si-Albert one
\cite{2,3}. Even more interesting is the observation of
non-trivial critical behaviour for these cases \cite{2,3,4,5}.

Another mechanism that leads to a finite transition temperature was discovered by one of us in the so called special attention network. This is a model of a weighted network where the interactions between connected spins are made topology-dependent \cite{6}. A further investigation of that and related models then led to the discovery that topology and interaction can be ´traded', in the sense that the effect of a change in interaction law can be transformed away by an appropriate change of the degree distribution law. In the present paper we present a detailed investigation of the critical behaviour of the Ising model with degree-dependent interactions. We also discuss the extension of our results to nonequilibrium situations. A summary of our results was published earlier \cite{0p}.

This paper is organised as follows. In the next section, we define the model and present the results of a simple mean field theory. In section 3, we study the model with the cavity approach. In section 4, we describe the application of the Bethe-Peierls method and the replica technique to our model. In section 5, we present the results of extensive Monte Carlo simulations. In section 6, we discuss the extension of our main result to a nonequilibrium process. Finally, we present our conclusions in section 7.

\section{The model}
\noindent
Our model can be defined on a general network (or graph).
When in a graph a vertex (or node) $i$ is connected with $k_i$ other nodes we say it has degree $k_i$. We will mostly have in mind a scale-free network for which the degree is a random variable whose distribution $P(k)$ is a power-law: $P(k) \propto k^{-\gamma}$. We take $\gamma > 2$ so that the average degree $Q=\int k P(k) dk$ is finite. For future reference we also introduce the notation $Q_2$ for the second moment of $P(k)$.
The Barab\'{a}si-Albert (BA) network has $\gamma=3$ \cite{1}. The number of nodes in the network is denoted by $N$.

We next define an Ising model on this network by associating to each node $i$ a variable $s_i=\pm 1$
and to each pair of linked nodes an energy $-J_{ij} s_i s_j$. In this paper, we will choose the couplings $J_{ij}$ to be given by
\begin{equation}
J_{ij} = J Q^{2\mu} /(k_i k_j)^\mu \,.
\label{1}
\end{equation}
with $J > 0$.
For $\mu=0$, the behaviour of this model has been investigated by various authors and the critical behaviour was found to depend on $\gamma$
\cite{2,3,4,5}. For $\gamma > 5$ the results of standard mean field theory were found to apply: the critical temperature $T_c$ is finite, the order parameter vanishes with an exponent $1/2$ and the specific heat has a finite jump at $T_c$. When $3 < \gamma < 5$, the transition temperature remains finite, the specific heat goes to zero continuously as a function of temperature with an exponent that depends on $\gamma$. The same is true for the order parameter. In the borderline case $\gamma=5$, logarithmic corrections appear. For all cases with $\gamma > 3$, the zero-field susceptibility diverges as $|T-T_c|^{-1}$.
Finally, when $2 < \gamma \leq 3$, the system is ordered at all temperatures.

When $\mu=1/2$, our model corresponds to the special attention
network (SAN) introduced earlier by one of us \cite{6}. A simple
mean field approach showed that on a BA network,  the SAN has a
finite transition temperature \cite{6}. We start by extending that
simple approach to the case of general $\mu$ and $\gamma$.

\subsection{A simple mean-field approximation}
\noindent
For a given network realization, the local magnetisation $m_i=\langle s_i \rangle $ at node $i$ obeys the exact equation
\begin{equation}
m_i=\langle s_i \rangle = \left\langle \tanh\left( \sum_{j=1}^{k_i} \frac{J_{ij}}{k_B T} s_j \right)\right\rangle \,.
\label{2}
\end{equation}
Here $\langle . \rangle$ is the thermal average, $k_B$ is Boltzmann's constant and $T$ is the temperature. Following Bianconi \cite{7}, we now apply a double mean field approximation to this equation by rewriting it as
\begin{equation}
m_i = \tanh \left( \sum_{j=1}^N \frac{[J_{ij}]}{k_B T} m_j\right) \, ,
\label{3}
\end{equation}
where the sum now runs over all the nodes and $[.]$ denotes the average over all the realisations of the network with a fixed set of degrees $\{ k_i \}$. For a BA network the probability $p_{ij}$ that two nodes are connected was shown to be equal to $k_i k_j /(QN)$, \cite{7} and we can expect this result to hold in some other cases \cite{5}.
We therefore have
\begin{equation}
[J_{ij}]=J_{ij} p_{ij} = \frac{J Q^{2\mu-1}}{N} \left( k_i k_j\right)^{1-\mu}\, ,
\label{4}
\end{equation}
so that Eq. (\ref{3}) can be rewritten as
\begin{equation}
m_i = \tanh \left( \frac{JQ^\mu}{k_B T}\  k_i^{1-\mu} \frac{Q^{\mu-1}}{N} \sum_{j=1}^N  k_j^{1-\mu} m_j \right) \, .
\label{5}
\end{equation}
The quantity
\begin{equation}
S = \frac{Q^{\mu-1}}{N} \sum_{j=1}^N  k_j^{1-\mu} m_j \, ,
\label{6}
\end{equation}
is a convenient order parameter.
From Eq. (\ref{5}) it follows that $S$ obeys the selfconsistency equation
\begin{equation}
S = \frac{Q^{\mu-1}}{N} \sum_{i=1}^N k_i^{1-\mu} \tanh\left( \frac{JQ^\mu}{k_B T} k_i^{1-\mu} S \right)
\label{7}
\end{equation}
For $N \to \infty$, the term on the right hand side is simply related to the average over the distribution $P(k)$ so that we can rewrite Eq. (\ref{7}) as
\begin{equation}
S = Q^{\mu-1} \int_m^{\infty} k^{1-\mu} \tanh\left(\frac{JQ^\mu}{k_B T} k^{1-\mu} S \right) P(k) dk
\label{8}
\end{equation}
Here $m \geq 1$ is the lowest degree that is possible in the network.
Eq. (\ref{8}) can be analysed in a standard way. For example, the critical temperature is determined by assuming $S$ to be small. After linearisation we then obtain
\begin{equation}
T_c = \frac{JQ^{2\mu-1}}{k_B} \int_m^\infty k^{2-2\mu} P(k) dk \,.
\label{9}
\end{equation}
For the power-law distribution this gives a finite $T_c$ provided $\gamma > 3-2\mu$. This is consistent with the earlier finding of a finite $T_c$ for the SAN on the BA network. One could then go on and determine, for example, the exponent $\beta$ from a further analysis of Eq. (\ref{8}). It is however more suitable to perform the transformation of variables
\begin{equation}
k' = Q^{\mu} k^{1-\mu}
\label{10}
\end{equation}
This gives for the case that the degree distribution is power-law, and for $\mu < 1$
\begin{equation}
S= A \int_{m'}^\infty \left(k'\right)^{\frac{1-\gamma}{1-\mu}} \tanh\left(\frac{Jk'S}{k_B T}\right) dk'
\label{11}
\end{equation}
where $A$ is a constant and $m'=Q^\mu m^{1-\mu}$. Comparison with (\ref{8}) teaches us that (\ref{11}) is precisely the mean field equation for $\mu'=0$ and for a degree distribution with an exponent that is modified to
\begin{equation}
\gamma' = \frac{\gamma-\mu}{1-\mu}
\label{12}
\end{equation}
This relation can be expected to hold more generally. Indeed, it should be valid whenever within a mean-field approach the degree $k$ only enters in physical properties through the quenched average interaction between any two nodes with fixed degrees, $k_i$ and $k_j$. This average is, as shown above, $J_{ij} p_{ij}$ with $p_{ij}=k_ik_j/(QN)$. For $\mu \neq 0$, the $k_i$ can then be transformed using (\ref{10}). In order to retain the same physics, averages over the degree distribution must be invariant. This requires a distribution transformation
\begin{eqnarray}
P(k)=P'(k'(k))\frac{dk'(k)}{dk}
\label{12accent}
\end{eqnarray}
from which (\ref{12}) follows for a scale-free $P(k)$. In section 6, we will in fact show that (\ref{12}) also holds for a contact process with degree-dependent infection rates. The general derivation scheme outlined above, will also apply there.
Thus, our simple mean field analysis indicates that by tuning $\mu \in [2-\gamma,1]$ we can encounter the whole range of universality classes that was found in earlier work at $\mu=0$.
As an example, one can study the whole set of universality classes by working on a BA-network
and tuning $\mu$. For the particular example of the SAN, (\ref{12}) gives $\gamma'=5$.

In the rest of this paper we will apply several other approaches that go beyond simple mean field to our model. These will allow us to get a better estimate than (\ref{9}) for the critical temperature, and to verify the equivalence between universality classes as expressed in Eq. (\ref{12}).

We also observe that from (\ref{5}) and (\ref{6}) it follows that
\begin{eqnarray}
m_i = \tanh \left(\frac{JQ^\mu}{k_B T} k_i^{1-\mu} S\right) \approx \frac{JQ^\mu}{k_B T} k_i^{1-\mu} S
\label{301}
\end{eqnarray}
where the last approximation holds close to $T_c$. We therefore anticipate that the local magnetisation is proportional to $k_i^{1-\mu}$, a result that we will use further in this paper and which will be verified numerically.

\section{The cavity method}
\noindent
We begin with a study of our model using the cavity approach. One reason for this is that concepts like the cavity field
and the propagated field, which also appear in the replica study of our model, can be introduced in a physically transparent way within the context of this method.

The cavity method is closely related to the well known Bethe-Peierls (BP) approximation for a spin model on a Cayley tree \cite{11}. In fact, both methods are equivalent when there is replica symmetry (for the precise relation between various mean-field approximations, see reference 20).
The cavity method was applied to diluted
spin models in references 21-23. For Ising like models on a network, the method has the advantage that it can take into account site degree
correlations, which certainly are present in networks grown via a preferential attachment rule \cite{12,13} or in real world networks (from biology, sociology, information technology, ...).
These correlations can be measured in terms of the clustering coefficient, the betweenness \cite{-2} or the abundance of cycles of a given
length \cite{14,15}.
This has to be contrasted with the replica approach of the next section, which, in its simplest form assumes independence of the site degrees.
Extensions of the replica approach that take into account degree correlations are known \cite{13}, but necessarily are more involved.
Another attractive property of the cavity method is that thermodynamic quantities like the magnetisation can be calculated on a single
network realisation. Together these properties allow the analysis of real world networks and to work with ensembles that, as in the BA case, are defined only through a growth process.


To derive the cavity equations, it is assumed that the network has the structure of a tree. It is at first sight paradoxical that the tree approximation is adequate
for determining the critical point of the network, because the Ising model
on a tree (without loops) cannot display spontaneous symmetry breaking (SSB)
at finite temperature and thus $T_c = 0$. However, in the same way that the
Bethe-Peierls approximation for an Ising model on a Cayley tree introduces
an effective symmetry breaking field in the bulk so that SSB becomes
possible, the cavity method  introduces a random distribution of effective
fields in the bulk (all fields being of the same sign) so that SSB becomes
possible notwithstanding the absence of loops on the tree. Thus, the SSB due
to the sparse loops in the actual network is replaced, in the cavity method,
by the SSB due to the effective fields. The question remains whether the two
SSB mechanisms lead to exactly the same value of $T_c$.

\begin{figure}[htbp]
\includegraphics[width=2.5in]{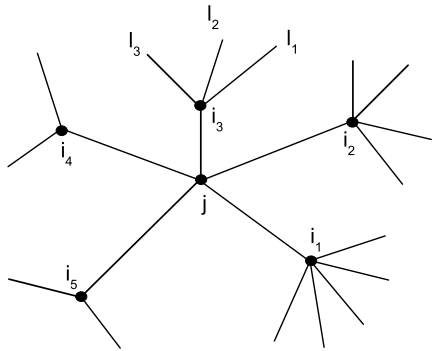}
\caption{Local structure of a network used in the derivation of
the cavity equations (see text).} \label{fig1}
\end{figure}

Consider then a particular site $j$ in the network (see Fig. \ref{fig1}).
The assumption of tree structure implies that the sites connected to $j$ form the roots of a set of independent subgraphs.

The site $j$ is connected to $k_j$ others. In the presence of an external field $H$ the magnetisation at site $j$ is given by (with $\beta=1/k_B T$)
\begin{eqnarray}
m_j = \frac{e^{\beta H} \prod_{i=1}^{k_j} Z_{ij}(+) - e^{-\beta H} \prod_{i=1}^{k_j} Z_{ij}(-)}
{e^{\beta H} \prod_{i=1}^{k_j} Z_{ij}(+) + e^{-\beta H} \prod_{i=1}^{k_j} Z_{ij}(-)}
\label{36}
\end{eqnarray}
Here $Z_{ij}(s)$ is the partition sum of the whole subgraph
starting from site $i$, and $s$ is the value of the spin at $j$.
This partition sum also includes the bond between the vertices $i$
and $j$. Clearly, we can always write
\begin{eqnarray}
Z_{ij}(s) \propto e^{\beta u_{ij} s}
\label{37}
\end{eqnarray}
where we call $u_{ij}$ {\it the local cavity message}.
With this assumption, Eq. (\ref{36}) becomes
\begin{eqnarray}
m_j = \tanh\left(\beta H + \beta \sum_{i=1}^{k_j} u_{ij}\right)
\label{38}
\end{eqnarray}
which also gives a clear physical interpretation to $u_{ij}$. Indeed, $u_{ij}$ can be seen as the contribution to the total magnetic field acting on site $j$ coming from the independent
subtree having site $i$ as root. In the cavity ansatz, those local contributions are indeed independent.
We need extra equations to determine the cavity fields selfconsistently. In order to determine these, we write $Z_{ij}(s)$
in terms of the subtrees that are connected to it (see Fig. \ref{fig1}). We denote the sites adjacent to $i$ by the index $l$. Their number equals $k_i$, but clearly, one of them is the starting site $j$. From the definition of $Z_{ij}(s)$, we have
\begin{eqnarray}
Z_{ij}(s) = e^{\beta H + \beta J_{ij}s}\prod_{l=1}^{k_i - 1} Z_{li}(+) + e^{-\beta H - \beta J_{ij}s}\prod_{l=1}^{k_i-1} Z_{li}(-)
\label{39}
\end{eqnarray}
so that using Eq. (\ref{36}) and Eq. (\ref{37}), we obtain after a little algebra
\begin{eqnarray}
u_{ij} = \frac{1}{\beta} \tanh^{-1} \left(\tanh\left(\beta h_{ij}\right)\tanh\left(\beta J_{ij}\right)\right)
\label{41}
\end{eqnarray}
where
\begin{eqnarray}
h_{ij}=H+ \sum_{l=1}^{k_i-1} u_{li}
\label{42}
\end{eqnarray}
is called {\it the propagated field} (sometimes also called cavity field). The equations (\ref{41}) -(\ref{42}) are known as belief propagation equations \cite{11}. They were derived iteratively here, but it is possible to obtain the same equations through a variational procedure and assuming that
configuration probabilities properly factorize on a tree. A short derivation from the variational approach
will be presented in section 4.1.

We solved the belief propagation equations iteratively for different values of $\mu$ in our model.
One can show in this case that belief propagation equations, when converging, do so to a unique fixed point (beside the
fully paramagnetic solution which is always present in absence of external field). When possible, we could also
be interested in averaging belief propagation equations over a proper random network ensemble. In that case, equations
for propagated messages and fields become a set of integral equations for field probability distributions \cite{9,10}. This unfortunately
cannot be easily done in the case of growing network ensembles, like the BA one. In the case of uncorrelated power-law degree distributed graphs, the self consistent integral equations can be written straightforwardly
from (\ref{41}) and (\ref{42}). The resulting equations turn out to be exactly equations (\ref{28}) and (\ref{29}) that we will obtain through the replica approach.
On the same random network ensemble, the replica symmetric
calculation and the cavity method are therefore completely equivalent. Note, however, that the analytical
results for $T_c$ presented in the replica calculation
are valid for a graph ensemble which is not exactly the
BA one. It is therefore plausible that we obtain a slight discrepancy between the replica and the cavity results
obtained after averaging over real BA network realisations.


Within the cavity approach, it is also possible to obtain the free energy $F(\beta)$ and from this other thermodynamic quantities such as the energy
$U(\beta)$ and the specific heat $C(\beta)$. Here we only quote the result for $U$
\begin{eqnarray}
U(\beta)= -  \sum_{i,j} J_{ij} \left( \frac{\tanh(\beta J_{ij}) + \tanh(\beta h_{ij})\tanh(\beta h_{ji})}{1+\tanh(\beta J_{ij})\tanh(\beta h_{ij})\tanh(\beta h_{ji})}\right)
\label{43}
\end{eqnarray}

We now turn to a discussion of our numerical results. These were obtained on BA networks.
Ensemble averages in the case of BA
networks can be performed numerically generating a large number of graphs with a given number of nodes with the usual preferential attachment
rule, and subsequently averaging the results over all the graphs at a given $N$. Quantities such as $T_c$ and the critical exponents can then
be found using finite size scaling. We studied $1000$ realisations of networks of $N=50,100,250,500,1000,5000$ and
$10^4$ nodes, $100$ realisations with $N=5\times 10^4$ nodes and $10$ realisations of $10^5$ and
$10^6$ nodes. For each of these the cavity messages were determined iteratively.
The numerical results showed little fluctuations when comparing different realisations, even for small network sizes.

We first investigated the SAN and studied cases with $Q=4,6,8,10$ and $20$.
In table 1 (see section 4) we give our estimates for $T_c$ as a function of $Q$.
These results will be compared with those coming from other methods.

\begin{figure}[htbp]
\centerline{\includegraphics[width=2.8in]{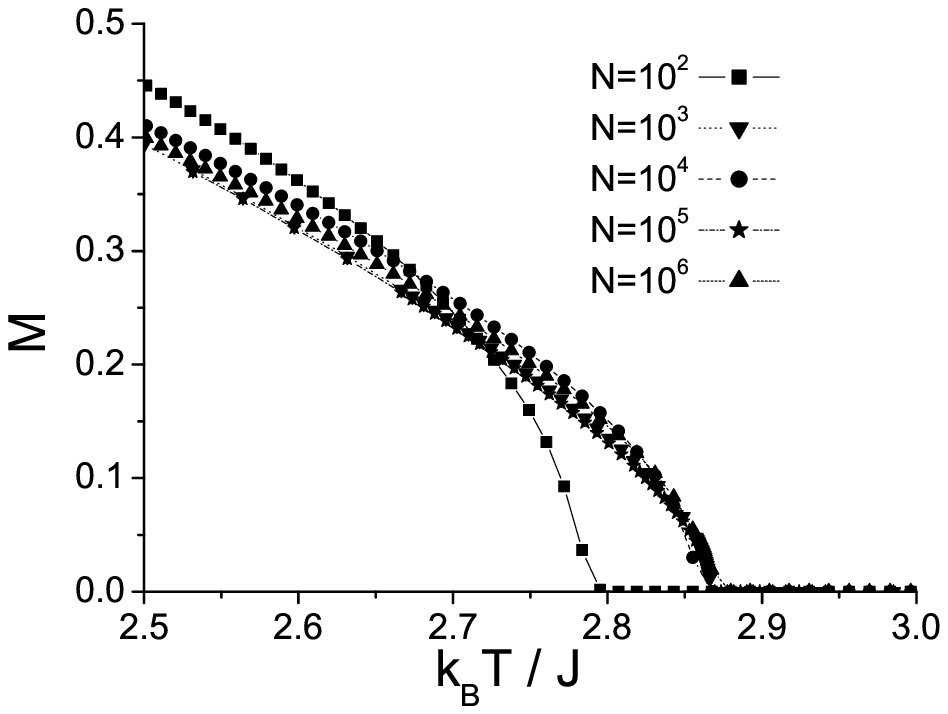}
\includegraphics[width=2.5in]{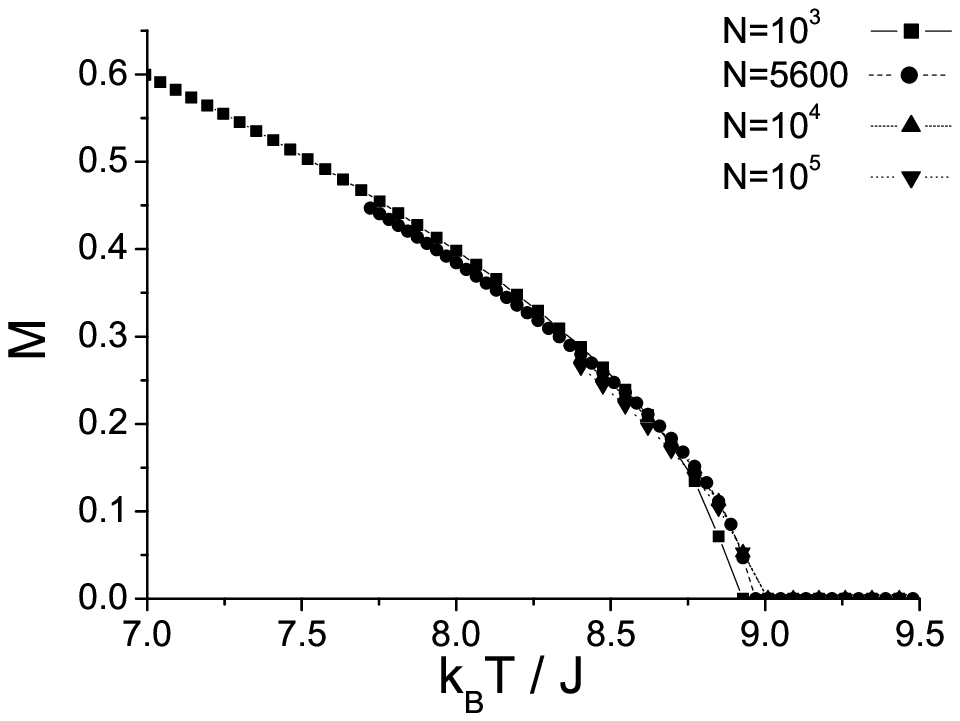}}
\caption{Magnetisation versus temperature for $Q=4$ (left) and
$Q=10$ (right) for $N=10^2, 10^3, 10^4, 10^5$. For $Q=4$, results
for $N=10^6$ are also included.} \label{fig2}
\end{figure}

From the behaviour of the magnetisation $M$ ($M=\frac{1}{N}\sum_i
m_i$) below $T_c$ (Fig. \ref{fig2}), we can determine the exponent
$\beta$. We find a value which is very close to the mean field
value of $1/2$. This value hardly depends on $N$ or on $Q$. These
data also allow us to obtain finite size estimates $T_c(N)$ for
the critical temperature. We measured the magnetisation within a
small temperature window around each value of $T_c(N)$ and this
for small values of the external field $H$. From an extrapolation
for large $N$, the value of the exponent $\delta$ was found to be
close to $3$. More precisely we find $1/\delta=0.333 \pm 0.010$
for $Q=4$. For bigger $Q$-values, we find the same value for
$1/\delta$ while the error decreases with $Q$. Using the scaling
relation $\gamma_s=\beta(\delta-1)$, this leads to the value $1$
for the susceptibility exponent $\gamma_s$.

\begin{figure}[htbp]
\centerline{\includegraphics[width=2.5in]{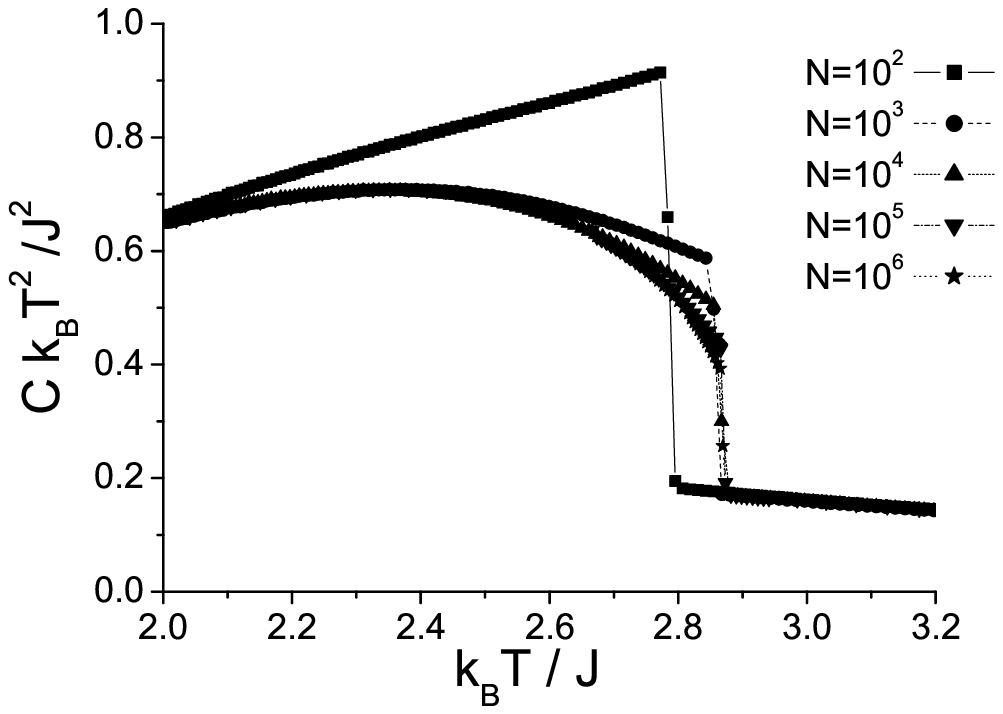}
\includegraphics[width=2.5in]{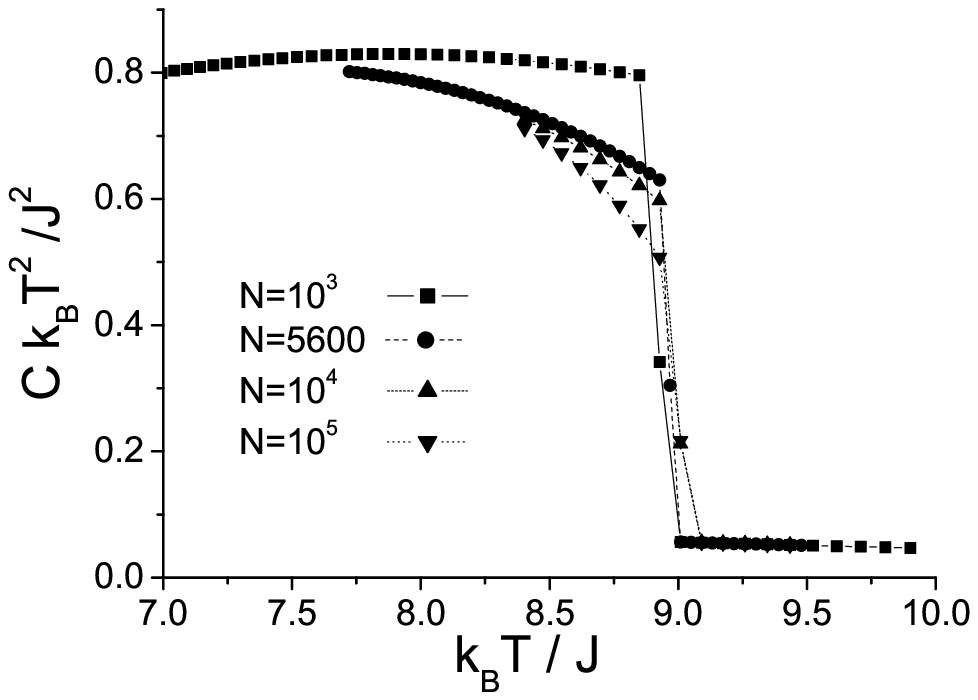}}
 \caption{Specific heat versus temperature for
$Q=4$ (left) and $Q=10$ (right) for $N=10^2, 10^3, 10^4, 10^5$.
For $Q=4$, results for $N=10^6$ are also included.} \label{fig3}
\end{figure}

Specific heat computations (see Fig. \ref{fig3}) show a jump around the transition. However, the discontinuity seems to become smaller for large $N$ such that
\begin{eqnarray}
\lim_{T \to T_c^-} \frac{dC}{dT} = - \infty
\label{44}
\end{eqnarray}
Both cases (vanishing jump with diverging derivative or discontinuity in $C$) can be consistent with a critical exponent $\alpha=0$.

The conjecture embodied in (\ref{12}) predicts that the SAN on a BA network has $\gamma'=5$. The critical exponents are then given by $\gamma_s=1$, $\beta=1/2$ and $\alpha=0$ \cite{2,3,4}.
In conclusion, we can say that the results from the cavity method for the SAN are in agreement with the relation (\ref{12}).

\begin{figure}[htbp]
\centerline{\includegraphics[width=2.5in]{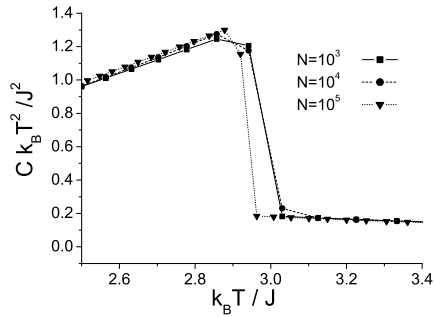}}
\caption{Specific heat versus temperature for a graph with a
degree distribution that is Poisson ($Q \approx 4$). The plots
refer to an average over 500 realisations for $N=10^3, 10^4$ and
50 realisations for $N=10^5$.} \label{fig4}
\end{figure}

A peculiar behaviour of the specific heat can be seen in Fig. \ref{fig3}. For both $Q$-values shown there, one observes a maximum in the specific heat
at a temperature below the critical one. This seems not to be a finite size effect. Moreover, as Fig. \ref{fig4} shows, this kind of behaviour does
not appear for the SAN on a network with a Poisson degree distribution.

\begin{figure}[htbp]
\centerline{\includegraphics[width=2.5in]{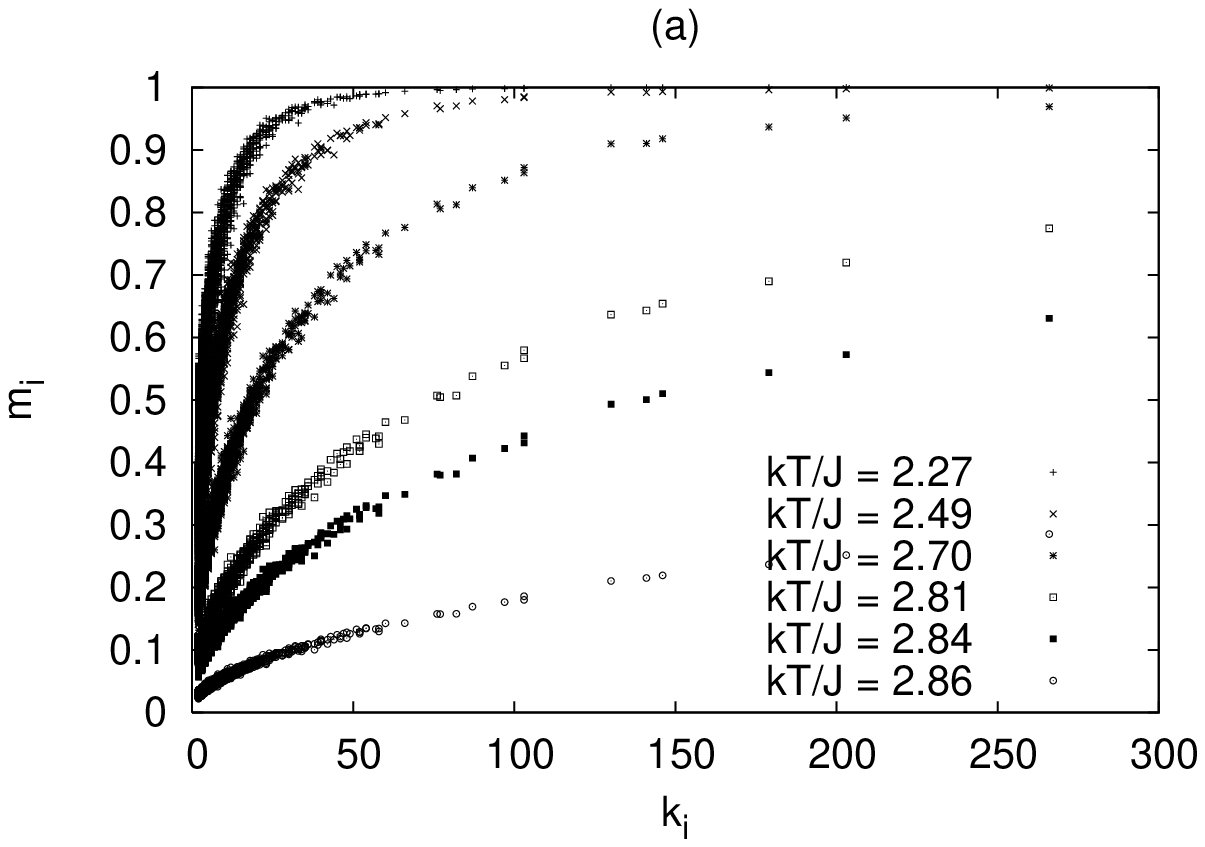}
\includegraphics[width=2.5in]{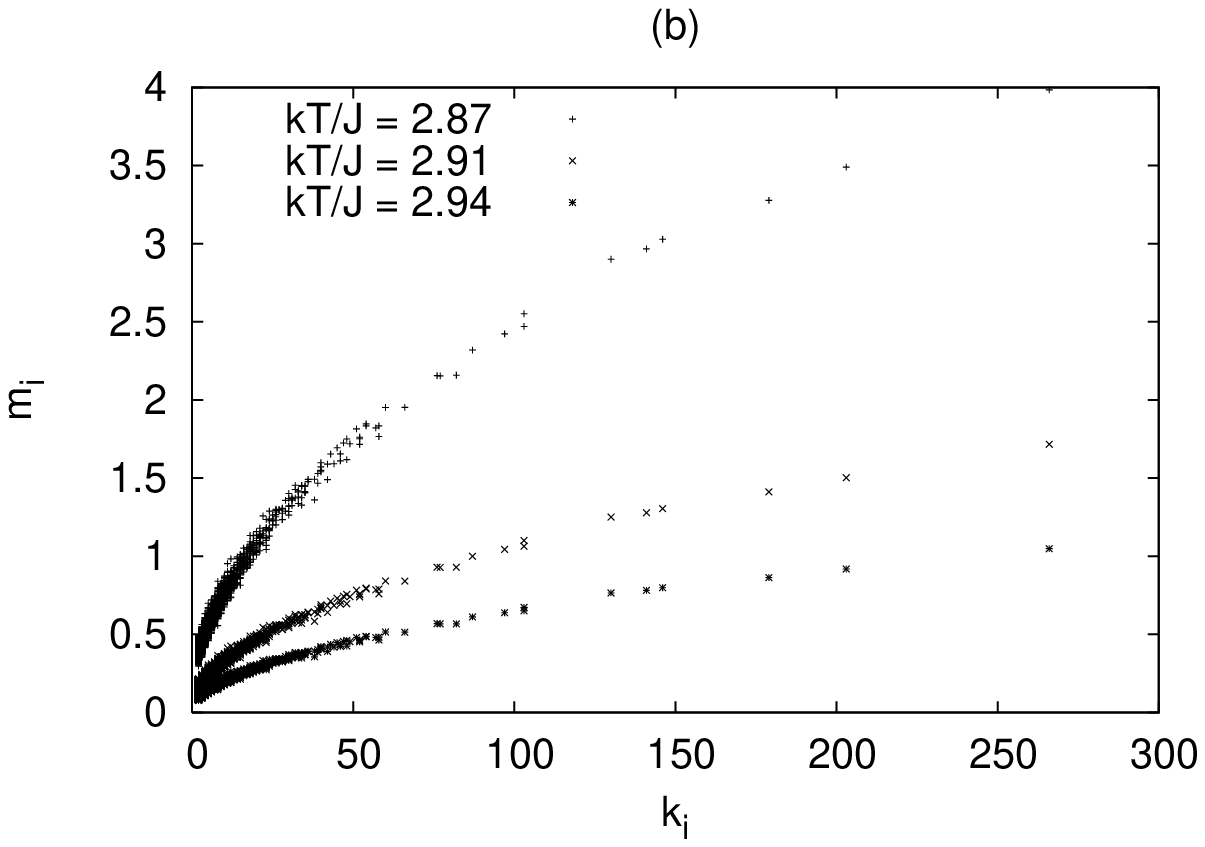}}
\caption{Scatter plot of single node magnetisations versus degree
for one network of $N=10000$ and $Q=4$ at different temperatures:
(a) below $T_c (k_BT_c/J=2.87\pm 0.01)$, (b) above $T_c$.}
\label{fig5}
\end{figure}

Another interesting feature is that close to $T_c$ we observe an inversion phenomenon in the local magnetisation of sites. Below the transition, there is a clear hierarchy in sites according
to their spin magnetisation. Even though couplings are weighted, hubs are still the most magnetized sites and are believed to drive the transition.
Plots of the local magnetization versus site degree for a few values of the temperature above and below $T_c$ are shown in Fig. \ref{fig5}. These data are consistent with the prediction (\ref{301}) coming from the simple mean field theory.

\begin{figure}[htbp]
\centerline{\includegraphics[width=2.5in]{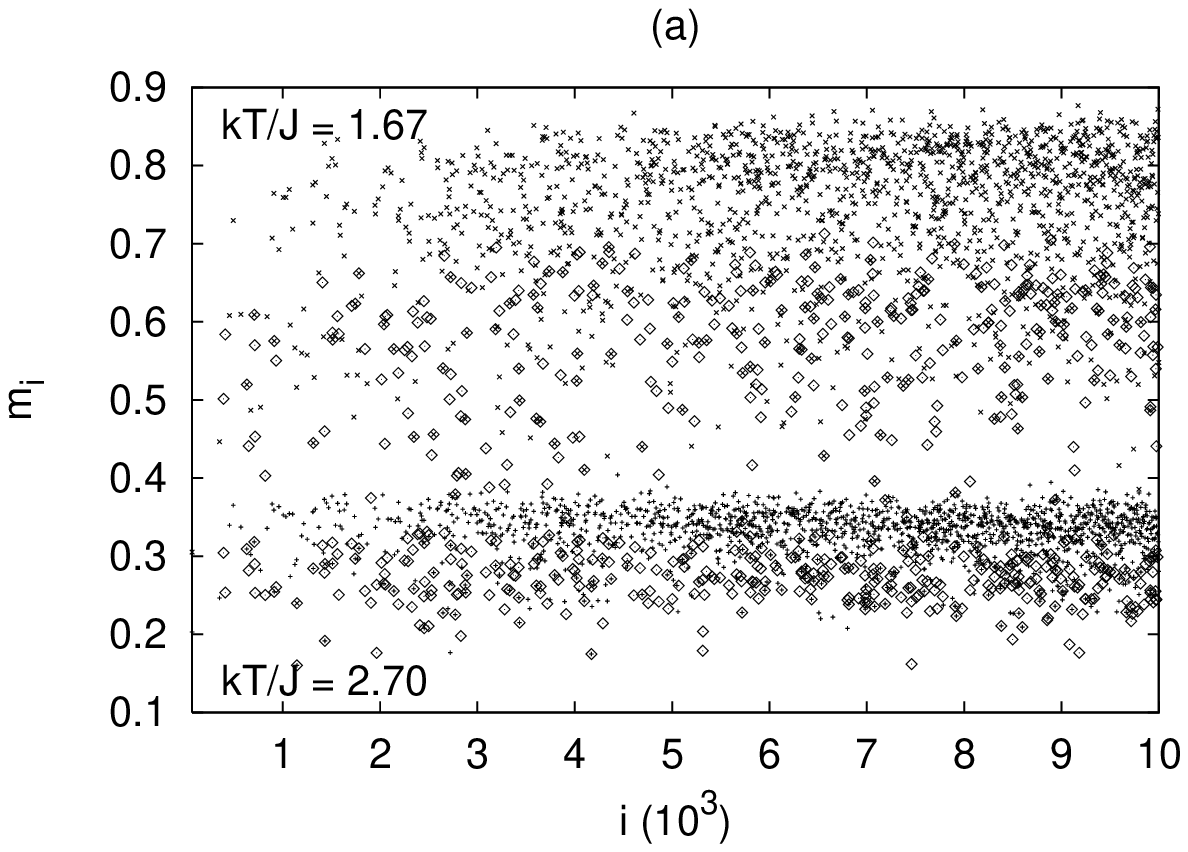}
\includegraphics[width=2.5in]{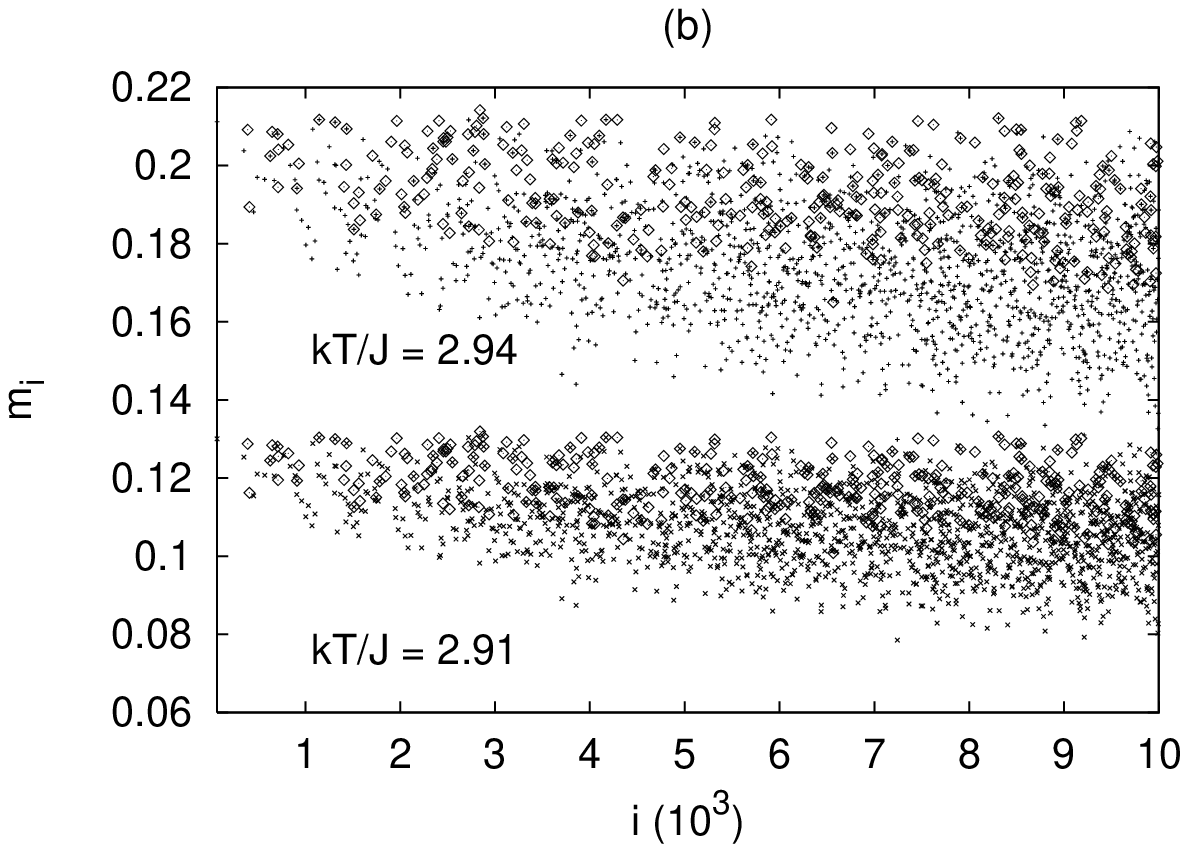}}
\caption{Scatter plot of magnetisations of all nodes of degree
$k=2$ for the same network as in Fig. \ref{fig5}, versus node
index (i.e. the time step at which this node is added when the
network grows according to the BA-rule). Diamonds represent those
nodes connected to hubs whereas crosses represent nodes of degree
2 not connected to hubs. Older nodes (smaller index) are on
average more connected to hubs than later ones. Different clusters
represent different temperatures. In (a) we show the results for
two temperatures below $T_c$ while in (b) the temperatures are
above the critical one. The inversion phenomenon described in the
text can clearly be seen. } \label{fig6}
\end{figure}

As can be seen in the scatter plots of Fig. \ref{fig6}, low degree
nodes connected to hubs have a lower magnetisation than equal
degree nodes that are not connected to hubs. Above the transition,
however, this situation is reversed. This is different from the
situation for an Ising model with degree-independent couplings on
a network, where at fixed degree, nodes connected to hubs are
always most magnetised.

\begin{figure}[htbp]
\centerline{\includegraphics[width=2.5in]{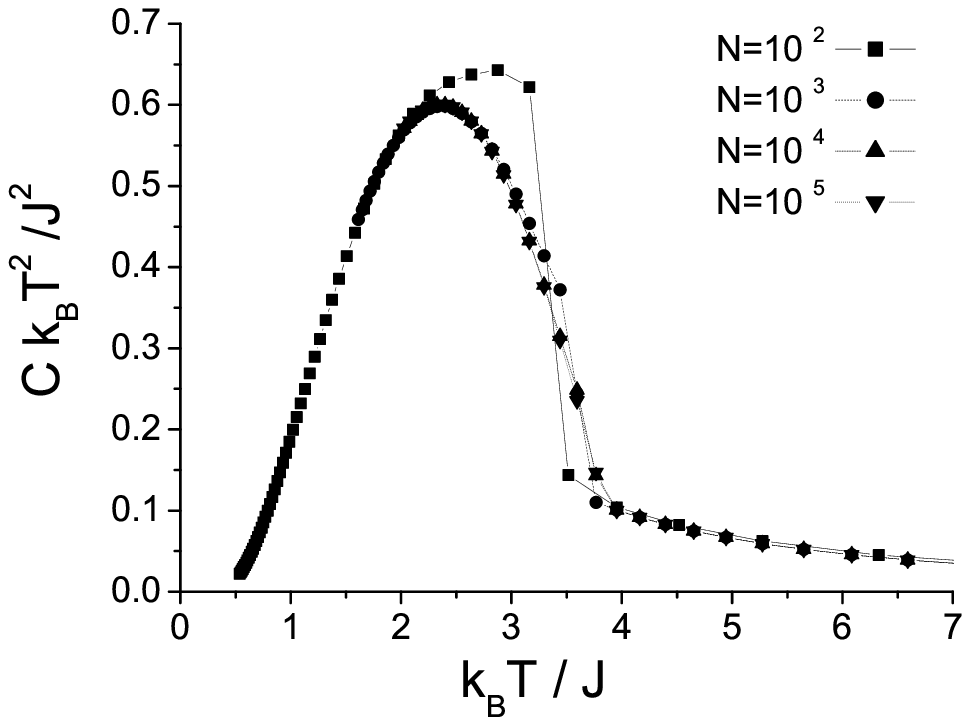}
\includegraphics[width=2.5in]{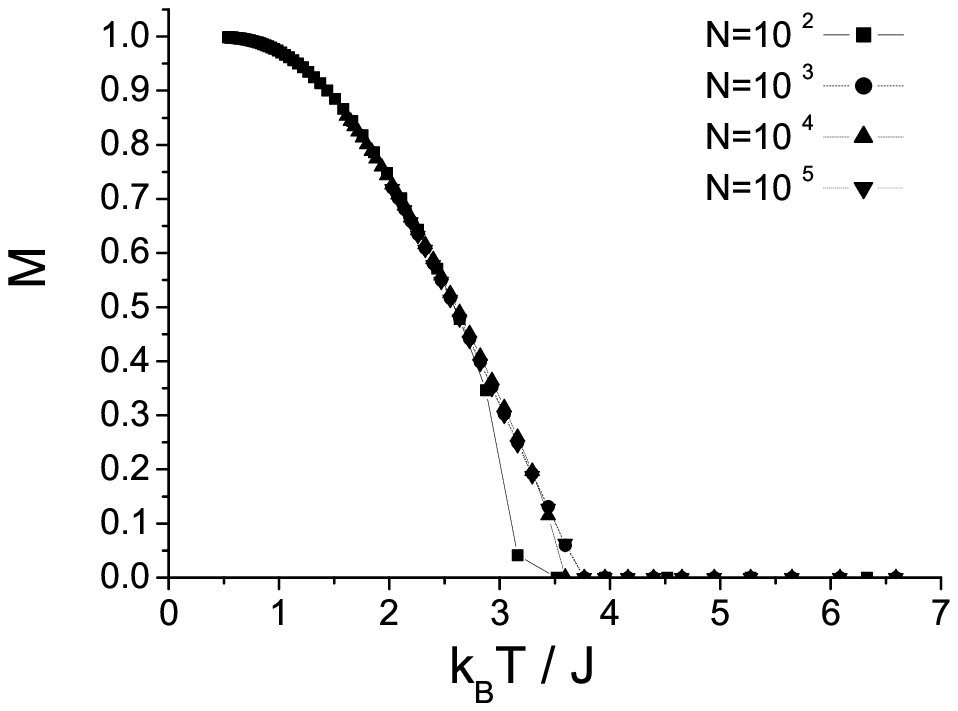}}
\caption{The specific heat (left) and the magnetisation (right) as
a function of temperature for a BA network with $\mu=1/3$. The
average connectivity is $Q=4$.  } \label{fig14}
\end{figure}

Besides the SAN, we also investigated our model with the same techniques for $\mu=1/3$ on a BA network. For that case, (\ref{12}) predicts
$\gamma'=4$ which leads to the exponent values $\gamma_s=\beta=-\alpha=1$ and $\delta=2$. In
Fig. \ref{fig14} we show some representative results for the specific heat and the magnetisation as a function of temperature (both at $Q=4$). They show the appearance with increasing $N$ of a  regime where both quantities depend linearly on temperature, consistent with the above prediction.

\section{The critical temperature}
In this section we discuss two approaches, the Bethe-Peierls approximation and the replica method, that allow us to get precise results for the critical temperature. Both approaches assume that there are no degree correlations present in the network.

\subsection{Bethe-Peierls approximation}
The Bethe-Peierls (BP) approximation is the simplest of the cluster variation methods \cite{16}.
It amounts to approximating the entropic part of the free energy, $F(\beta)$, by restricting the probability distribution $\rho$ of a configuration of $N$ spins to a combination of single-site and nearest-neighbour pair distributions. For a particular fixed network, this leads to the following expression for $F(\beta)$
\begin{eqnarray}
F(\beta)\beta \approx \beta \langle H \rangle  + \sum_i (1-k_i) \sum_{s_i=\pm 1}
\rho_i \ln \rho_i + \frac{1}{2} \sum_i \sum_{j=1}^{k_i} \sum_{s_i,s_j=\pm 1}
\rho_{ij} \ln \rho_{ij}
\label{201}
\end{eqnarray}
where
\begin{eqnarray}
H &=& - \frac{1}{2} \sum_i \sum_{j=1}^{k_i} J_{ij} s_i s_j  \\
\rho_i &=& \frac{1 + m_i s_i}{2} \\
\rho_{ij} &=& \frac{1 + m_i s_i + m_j s_j + \psi_{ij} s_i s_j}{4}
\end{eqnarray}
Here $\psi_{ij}=\langle s_i s_j \rangle$. Expressions for the local magnetisations $m_i$ and the neighbour spin-spin correlations
$\psi_{ij}$ are obtained by looking for extrema of the free energy
\begin{eqnarray}
\frac{\partial F}{\partial m_i} &=& 0  \ \  \ \ \ \ \ \ \forall \ i \label{extr1}\\
\frac{\partial F}{\partial \psi_{ij}} &=& 0 \ \  \ \ \ \ \ \ \forall \ i,j \label{extr2}
\end{eqnarray}
It can be shown that (\ref{extr1}) and (\ref{extr2})
are fully equivalent to (\ref{41}) and (\ref{42}) \cite{11}.

Next, the resulting selfconsistency equations are linearised in the $m_i$. In this way one obtains
for the spin-spin correlations
\begin{eqnarray}
\psi_{ij} = \tanh {(J_{ij}/k_B T)}
\label{201a}
\end{eqnarray}
After summing over all the vertices, the equation for the magnetisation becomes
\begin{eqnarray}
\frac{1}{N} \sum_r (k_r - 1) m_r = \frac{1}{N} \sum_r m_r \sum_{j=1}^{k_r}\frac{1}{1+\psi_{rj}}
\label{202}
\end{eqnarray}
Since we now want to get a simpler {\it analytic} expression for the critical temperature,
we do not solve (\ref{202}) on single graphs, as already done in a more complete setting with the cavity approach.
Instead, we average it over the network realizations.
This we do in two steps.
Firstly, using a similar approximation
as in section 2.1, we replace the second sum on the right hand side by a sum over all nodes. Hence, we obtain
\begin{eqnarray}
\frac{1}{N} \sum_r (k_r - 1) m_r = \frac{1}{N} \sum_r m_r \sum_{j=1}^N p_{rj} \left( \frac{1}{1+\psi_{rj}}\right)
\end{eqnarray}
where again $p_{rj}=k_r k_j/(QN)$. Secondly, we insert the proportionality between $m_j$ and $k_j^{1-\mu}$ implied by (\ref{301}): $m_j \propto k_j^{1-\mu}$. In the limit $N \to \infty$ the resulting equation can again be written in terms of the average over the degree distribution. We then finally obtain
\begin{eqnarray}
Q \sum_{k=m}^\infty P(k) (k-1) k^{1-\mu}  = \sum_{k_1=m}^\infty P(k_1) k_1^{2-\mu} \sum_{k_2=m}^\infty P(k_2)
\frac{k_2}{1+\tanh\left(\frac{JQ^{2\mu}}{k_BT k_1^\mu k_2^\mu}\right)}
\label{305}
\end{eqnarray}
For the case $\mu=0$, and for $Q$ large, (\ref{305}) can be approximated and gives the estimate $k_B T_c/J \approx ( Q_2 - Q)/Q $, in agreement with exact results \cite{2}. Also, for the SAN on a BA-network, we get for $Q$ large: $k_B T_c/J \approx Q-1$. On a regular lattice with coordination number $Q$, the Bethe-Peierls approximation gives $J/(k_BT_c)=\ln{\left(\frac{Q}{Q-2}\right)}/2$ \cite{777}. As can be seen in table 1, this approximation gives a very good value for $T_c$ for $Q>4$.

Equation (\ref{305}) can be solved numerically. In table 1 we present our results for the SAN on a BA network with $P(k)=2 m (m+1) /k(k+1)(k+2)$ (for this case, $m=Q/2$).

\begin{table}[htbp]
\caption{Critical temperature $k_BT_c/J$ as a function of $Q$ for
the SAN on networks, as obtained from various approaches. On a
regular lattice with coordination number $Q$, the Bethe-Peierls
(BP) approximation gives $J/k_BT_c=\ln\left(Q/(Q-2)\right)/2$.
This value is given in the second column. In the third column the
results of the cavity approach are given, while the fourth column
presents the estimate obtained within the Bethe-Peierls
approximation on the network, equation (\ref{305}). The fifth
column gives the solution to (\ref{35}) obtained in the replica
approach. The last column gives the estimates coming from the
Monte Carlo simulations. The results for the cavity method and the
simulations were obtained on BA networks. The results for the
network BP and the replica method were obtained on uncorrelated
networks. This has dramatic consequences for $Q=2$. For $Q=2,
T_c=0$ on BA networks because BA networks with $Q=2$ are simple
trees without loops. In contrast, uncorrelated networks with $Q=2$
may feature loops as well as disconnected parts.} \centerline{
\begin{tabular}{r c c c c c}\\
\hline
$Q$ & lattice BP  &cavity & network BP &
replica &Monte Carlo\\
\hline
2 & 0.& 0. &  0.95230  & 0.94614 & 0. \\
4 & 2.8854 & 2.87$\pm$ 0.01 &2.95508 & 2.92468&2.91 $\pm$ 0.02\\
6 & 4.9326 & 4.92$\pm$ 0.01 &4.9554 &  4.94511 &{}\\
8 & 6.9521 & 6.94$\pm$ 0.01 &6.9406 &6.95796 &\\
10 & 8.9628 & 8.94$\pm$ 0.01 &8.9087 &8.96607 &8.96 $\pm$ 0.01\\
20 & 18.9824 & 18.98 $\pm$ 0.01 & 18.9107 &  18.98187 &19.09 $\pm$ 0.04 \\
\hline\\
\end{tabular}}
\end{table}

\subsection{Replica theory for an uncorrelated network}
\noindent
The replica approach is a powerful mathematical technique that provides a way to perform an average over random network ensembles. For the present model, and in absence of degree correlations, it leads to a relatively simple analytical equation from which the critical temperature can be determined numerically.

In order to define the random ensemble of networks we start by defining the adjacency matrix $C$ of a graph.
This matrix is the $N \times N$ matrix whose element $c_{ij}$ is one if there is a link between node $i$ and $j$ and zero otherwise. Clearly,
\begin{eqnarray}
\sum_{j=1}^N c_{ij}= k_i
\label{14}
\end{eqnarray}
Moreover, for undirected graphs $C$ is symmetric, $c_{ij}=c_{ji}$. Using the adjacency matrix we can write the Hamiltonian
${\cal H}$ for our model as
\begin{eqnarray}
{\cal H} = -\frac{1}{2} \sum_{i \neq j}^N J \Phi(k_i,k_j) c_{ij} s_i s_j - H \sum_{i=1}^N s_i
\label{15}
\end{eqnarray}
where in our case $\Phi(k_i,k_j)= J_{ij}/J= Q^{2\mu}/(k_i k_j)^{\mu}$. The replica approach can, however, be performed for more general forms of $\Phi$.
In Eq. (\ref{15}), we have also included an external
magnetic
field $H$ for later convenience.

We calculate the typical properties of all networks with a given fixed set of degrees $\{k_i\}$. The matrix elements $c_{ij}$ are assumed to be completely uncorrelated apart from the constraint (\ref{14}). Therefore given that
\begin{eqnarray}
P(c_{ij}) = \frac{Q}{N} \delta_{c_{ij},1} + \left(1-\frac{Q}{N}\right) \delta_{c_{ij},0}
\label{16}
\end{eqnarray}
we can write their joint probability distribution as
\begin{equation}
{\cal P}(\{c_{ij}\}) \propto \prod_{i<j} P(c_{ij}) \prod_i\delta\left(\sum_j c_{ij}-k_i\right)
\label{1600}
\end{equation}
As usual in disordered systems, we compute the quenched average of the free energy density per site, $f_q(\beta)$, from which typical properties can be determined
\begin{eqnarray}
-\beta f_q(\beta) = \lim_{N \to \infty} \frac{1}{N} \left[ \ln Z \right]
\label{17}
\end{eqnarray}
where $\beta=1/(k_B T)$.
Since the calculation is rather involved but standard \cite{3,8,17}, we summarize here only the major steps and results.
Using the equality $\ln Z = \lim_{n \to 0} \frac{1}{n} (Z^n-1)$ the quenched free energy density is written as
\begin{eqnarray}
-\beta f_q(\beta) = \lim_{N \to \infty} \lim_{n \to 0} \frac{1}{Nn}\left( \left[ Z^n (\beta)\right]-1\right)
\label{18}
\end{eqnarray}
The replicated partition function $Z^n(\beta)$ equals
\begin{eqnarray}
Z^n(\beta) = \sum_{\vec{s}_1} \cdots \sum_{\vec{s}_N} \exp\left[ \frac{\beta}{2} \sum_{\alpha=1}^n
\sum_{i,j} J_{ij} s_i^\alpha s_j^\alpha + \beta H \sum_{\alpha=1}^n \sum_{i=1}^N s_i^\alpha \right]
\label{19}
\end{eqnarray}
where $\vec{s}_i=\{s_i^1,\dots,s_i^n\}$.
For the distribution of connectivities given in Eq. (\ref{1600}), the average $\left[ \cdot \right]$ is
\begin{eqnarray}
\left[ A(\{c_{ij}\}) \right] = \frac{1}{{\cal N}} \int \left[\prod_{i<j} dc_{ij} P(c_{ij}) \right]
\prod_{i=1}^N \delta\left( \sum_j c_{ij}-k_i\right) A(\{c_{ij}\})
\label{20}
\end{eqnarray}
where ${\cal N}$ is the normalisation
\begin{eqnarray}
{\cal N} = \int \left[\prod_{i<j} dc_{ij} P(c_{ij}) \right]
\prod_{i=1}^N \delta\left( \sum_j c_{ij}-k_i\right)
\label{21}
\end{eqnarray}
In order to calculate this average, it is common to introduce an
exponential representation of the constraint
\begin{eqnarray}
\delta\left( \sum_{j=1}^N c_{ij} - k_i\right) = \int_0^{2\pi} \frac{d\xi_i}{2\pi} e^{i \xi_i \left( \sum_{j=1}^N c_{ij} - k_i \right)}
\label{22}
\end{eqnarray}
A straightforward but lengthy calculation in which we integrate over the disorder and the auxiliary variables $\xi_i$ allows us to
obtain the free energy density in terms of the functional order parameter
\begin{equation}
R_k(\vec{s})=\frac{1}{N}\sum_{i=1}^N\delta_{\vec{s},\vec{s}_i}\delta_{k,k_i}e^{i\xi_i}
\label{23}
\end{equation}
and its canonical conjugate $\hat{R}_k$. This order parameter is
an order parameter in the replica space which is the joint density
of finding a spin configuration $\vec{s}$ with average
connectivity $k$ at each site, when a link has been removed from
that site. The result is
\begin{eqnarray}
-\beta
f_q(\beta)=\mbox{Extr}_{\{\widehat{R}_k(\vec{s}),R_k(\vec{s})\}}\Bigg\{-Q
\sum_{\vec{s},k}\widehat{R}_k(\vec{s})R_k(\vec{s})+\frac{Q}{2}+\nonumber
\\
\sum_{k}P(k)\ln\sum_{\vec{s}}\big[\widehat{R}_{k}(\vec{s})\big]^{k}e^{\beta
H\sum_{\alpha=1}^ns^\alpha}
+\frac{Q}{2}\sum_{k,k'}\sum_{\vec{s},\vec{\sigma}}R_k(\vec{s})R_{k'}(\vec{\sigma})
e^{\beta J\Phi(k,k')\vec{s}\cdot\vec{\sigma} }\Bigg\} \label{24}
\end{eqnarray}
where we also performed the rescaling $i\widehat{R}_{k}(\vec{s})\to- Q \widehat{R}_{k}(\vec{s})$.

Stationarity of the free energy with respect to $R_k(\vec{s})$ and $\widehat{R}_k(\vec{s})$ leads to the saddle-point equations
\begin{eqnarray}
\widehat{R}_k(\vec{s}) &=& \sum_{k',\vec{\sigma}} R_k'(\vec{\sigma})e^{\beta J \Phi(k,k') \vec{s}\cdot \vec{\sigma}} \nonumber \\
R_k(\vec{s}) &=& \frac{kP(k)}{Q} \frac{\left[\widehat{R}_k(\vec{s})\right]^{k-1} e^{\beta H \sum_{\alpha=1}^n s^\alpha}}{\sum_{\vec{\sigma}}\left[\widehat{R}_k(\vec{\sigma})\right]^k e^{\beta H \sum_{\alpha=1}^n \sigma^\alpha}}
\label{25}
\end{eqnarray}
If all the $\Phi(k,k')$ are positive, so that the model only has ferromagnetic interactions, it can be expected that the solution to these
equations has replica symmetry (RS). For the order parameter and its conjugate, this assumption takes the form
\begin{eqnarray}
R_k(\vec{s}) = \int dh W_k(h) \frac{e^{\beta h \sum_{\alpha=1}^n s^\alpha}}{\left[2 \cosh(\beta h)\right]^n}
\label{26}
\end{eqnarray}
and
\begin{eqnarray}
\widehat{R}_k(\vec{s}) = \int du Q_k(u) \frac{e^{\beta u \sum_{\alpha=1}^n s^\alpha}}{\left[2 \cosh(\beta u)\right]^n}
\label{27}
\end{eqnarray}
respectively. Here $h$ and $u$ are the local cavity message and the propagated field that we already encountered in section 2.
While the cavity approach calculates these fields for a specific network, in the replica approach we can obtain their probability distributions
over the set of all networks obeying the constraints (\ref{14}). These distributions are denoted as $W_k(h)$ and $Q_k(u)$ respectively.
Within the RS ansatz, the saddle-point equations become
\begin{eqnarray}
Q_k(u)&=& \sum_{k'} \int dh W_{k'}(h) \delta\left(u - \frac{1}{\beta} \tanh^{-1}\left[\tanh(\beta h) \tanh\left[\beta J \Phi(k,k')\right]\right]\right) \label{28}\\
W_k(h)&=& \frac{P(k)k}{Q} \int\left[\prod_{l=1}^{k-1} du_l Q_k(u_l)\right]\delta\left(h-\sum_{l=1}^{k-1} u_l - H\right)
\label{29}
\end{eqnarray}
and the free energy density equals
\begin{eqnarray}
\beta f_q (\beta)&=& - Q \ln{2} + Q \sum_k \int dudhQ_k(u)W_k(h) \ln\left[1+\tanh(\beta u)\tanh(\beta h)\right] \nonumber \\
&-& \sum_k P(k) \int \left[\prod_{l=1}^k du_l Q_k(u_l)\right] \ln\left[\frac{2 \cosh\left[\beta\left(\sum_{l=1}^k u_l + H \right)\right]}{\prod_{l=1}^k 2 \cosh\left(\beta u_l \right)}\right] \label{30}\\
&-& \frac{Q}{2} \sum_{k,k'} \int dhdh' W_k(h)W_{k'}(h') F_{kk'}(\beta,h,h') \nonumber
\end{eqnarray}
where
\begin{eqnarray}
F_{kk'}(\beta,h,h')=\ln\left[\cosh\left[\beta J \Phi(k,k')\right]
+\sinh\left[\beta J \Phi(k,k')\right] \tanh(\beta h) \tanh(\beta
h')\right] \nonumber
\end{eqnarray}
The average zero-field magnetisation per site, $M$, then follows immediately
\begin{eqnarray}
M=-\frac{\partial f_q}{\partial H} (H=0) = \sum_k P(k) \int \left[\prod_{l=1}^k du_l Q_k(u_l)\right]
\tanh\left(\beta\sum_{l=1}^k u_l \right)
\label{31}
\end{eqnarray}
From Eq. (\ref{28}) and Eq. (\ref{29}) we can obtain an equation that contains only the functions
$Q_k(u)$
\begin{eqnarray}
Q_k(u)=\sum_q \frac{P(q)q}{Q}\int\left[\prod_{l=1}^{q-1} du_lQ_q(u_l)\right]\delta\left(u-G_{kq}(\beta)\right)
\label{32}
\end{eqnarray}
where
\begin{eqnarray}
G_{kq}(\beta)=\frac{1}{\beta}\tanh^{-1}\left[\tanh\left(\beta\sum_{l=1}^{q-1} u_l\right) \tanh\left[\beta J \Phi(k,q)\right]\right]
\label{32pr}
\end{eqnarray}

One can try to solve this set of equations, for example using population dynamics techniques. Here we only investigate the simpler question
 of locating the critical temperature. For this, we assume that the cavity fields $u$ are very small near the transition.
The argument of the delta-function in Eq. (\ref{32}) can then be linearised using
\begin{eqnarray}
G_{kq}(\beta) \approx \tanh\left[\beta J \Phi(k,q)\right] \sum_{l=1}^{q-1} u_l
\nonumber
\end{eqnarray}
We next multiply both sides in Eq. (\ref{32}) by $u$ and we integrate. This gives
\begin{eqnarray}
\int du Q_k(u) u = \sum_q \frac{P(q) q}{Q} \tanh\left[\beta J \Phi(k,q)\right]\sum_{l=1}^{q-1} \int du_l Q_q(u_l) u_l
\label{33}
\end{eqnarray}
Finally, we denote $x_k=\int Q_k(u)u du$ and define a matrix $A(\beta)$ with elements
\begin{eqnarray}
A_{q,k}(\beta)= \frac{P(q) q (q-1)}{Q} \tanh\left[\beta J \Phi(k,q)\right] \ \ \ \ q,k \geq m
\label{34}
\end{eqnarray}
($m$ is the smallest degree appearing in the network). Finding the
critical temperature now amounts to locating the value of $\beta$
for which the matrix $A$ has an eigenvalue $1$, i.e. to solving
\begin{eqnarray}
\det \left( A(\beta_c) - I \right) =0 \label{35}
\end{eqnarray}

In principle, the matrix $A$ is infinite dimensional. For a given
$P(k), m$ and $\Phi(k,k')$ one can, however, calculate an estimate
$T_c(K)$ for the critical temperature by truncating the matrix and
limiting $q$ and $k$ to be smaller than a given $K$. An
extrapolation for $K \to \infty$ then gives $T_c$. We have
performed such a calculation for the SAN for different values of
$Q$. Some numerical values can be found in table 1.

\section{Monte Carlo simulation results}
\noindent
Finally, we also investigated our model numerically on a BA network. This allows us to investigate
effects that are neglected within the cavity approach, such as the appearance of loops in the network.

Firstly, we investigated the SAN with various values of $Q$. When
a BA network is grown by the addition of $m$ new links, one always
ends up with an even value of $Q$, since $Q=2m$. In order to
obtain an odd average connectivity, alternating values of $m$ were
used. For example, by alternating $m$ between $1$ and $2$ we
obtain $Q=3$. We used the standard Metropolis algorithm to
simulate our model. Typically, we averaged over $400$ uncorrelated
spin configurations at each temperature and for each realisation
of the network.

The simulations were made for networks with $N$ ranging between $100$ and $56000$. At low values of $Q$ we found large fluctuations in
the properties of different realisations. We typically investigated between $200$ and $5000$ realisations of the network, depending on the size of the
network and the quantity under investigation.

\begin{figure}[htbp]
\centerline{\includegraphics[width=2.5in]{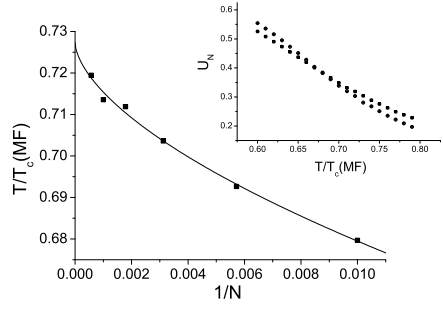}}
\caption{Finite-size estimate of the critical temperature (divided
by the mean field estimate $JQ/ k_B$) as a function of $1/N$
obtained using the intersections of the cumulants. An example of
such an intersection is given in the second plot. The average
connectivity is $Q=4$.} \label{fig7}
\end{figure}

The value of $T_c$ was obtained from the intersection of the Binder cumulants $U_N = 1 - \langle M^4 \rangle / (3 \langle M^2 \rangle^2)$, with $M$ the total magnetisation in a given configuration.
In order to obtain a value for the intersection point, we first fitted a curve through the data points (generally a fifth degree polynomial)
and then solved for the intersections of the fitted curves. From this, we obtain the finite-size ´critical temperature' $T_c(N)$.
These were then extrapolated using the standard relation
\begin{eqnarray}
T_c(N) \simeq T_c(\infty) - bN^{-\frac{1}{2-\alpha}\left(1 + \omega \nu\right)}
\label{45}
\end{eqnarray}
where $\omega$ is a correction-to-scaling exponent, introduced by
Binder \cite{201,202,203}, and $\nu$ the ``correlation length"
exponent proper to lattice models. In our network models the
product $\omega \nu$ figures as a single number, since a
correlation length is not defined. In Fig. \ref{fig7}, a typical
example of such a fit is shown. The resulting estimate of $T_c$,
together with those obtained for other $Q$-values, is given in
table 1. Moreover, from the data in Fig. \ref{fig7}, the estimate
$(1+\omega\nu)/(2-\alpha) =0.58\pm 0.12$ is obtained. The network
with $Q=3$ was only investigated with the simulation method and
for this case we find $k_B T_c/J =1.85 \pm 0.03$.

From the Binder cumulant, we can estimate the exponent $\alpha$. Indeed, the derivative of the cumulant $U'_N=dU_N/dT$ at $T_c(N)$ scales as
\begin{eqnarray}
\frac{U'_N (T_c)}{U'_{N'}(T_c)} = \left( \frac{N}{N'}\right)^{\frac{1}{2-\alpha}}
\label{46}
\end{eqnarray}
The derivative can be obtained from the polynomial fit used to determine $T_c(N)$. The values that we obtain are consistent with
the assumption that $\alpha=0$.

\begin{figure}[htbp]
\centerline{\includegraphics[width=2.5in]{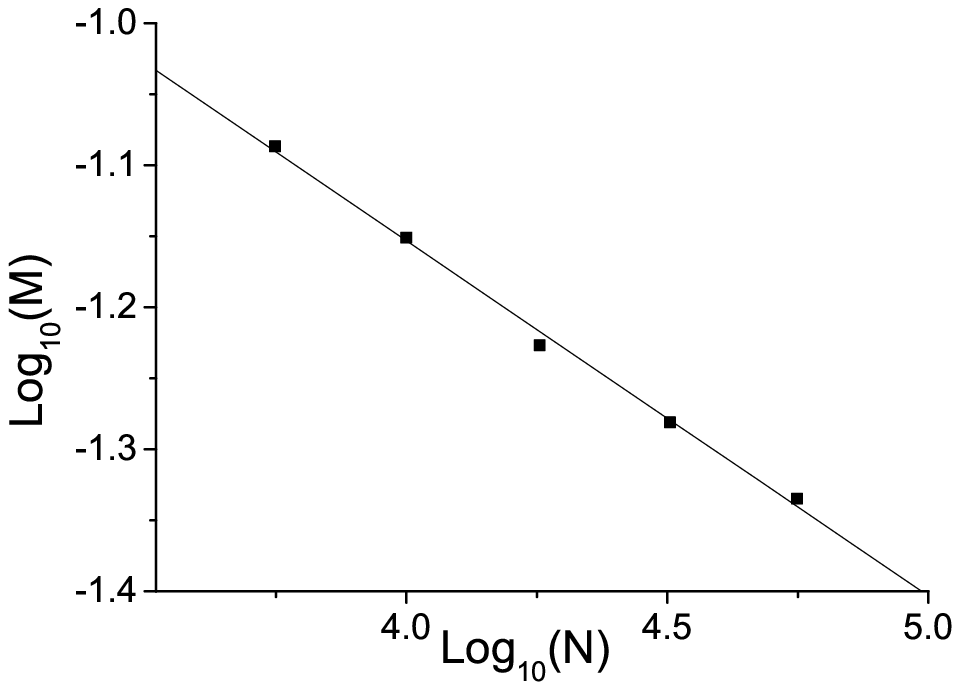}
\includegraphics[width=2.5in]{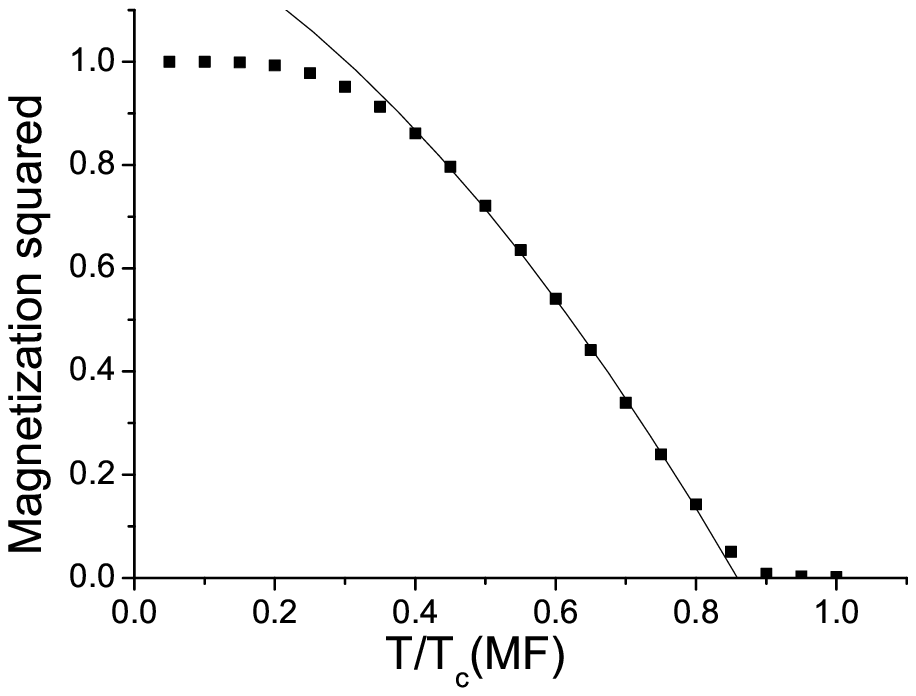}}
\caption{$N$-dependence of the magnetisation of the SAN at
$T_c(N)$ ($Q=10$). The slope is $-0.25\pm 0.01$. The right side
shows the magnetization squared as a function of temperature for a
network of 3200 nodes, averaged over 400 samples.} \label{fig8}
\end{figure}

Next, we calculated the magnetisation at $T_c(N)$ as a function of $N$. This allows us to estimate the exponent $\beta/(2-\alpha)$
as $\simeq 0.25$. In Fig. \ref{fig8} (left side) we show our numerical results for $Q=10$. This result is consistent with the mean field values $\beta=1/2$ and $\alpha=0$. As a further
test, we plotted the squared magnetisation as a function of $T$. Below $T_c$, we expect this to be a linear function and from a fit
to this form we can get a second estimate of $T_c$ (see Fig. \ref{fig8}, right side). The values that we find in this way are in agreement with those
coming from the cumulants, although the precision is less good.

We also computed the finite-lattice susceptibility $\chi'$, introduced by Binder \cite{18}
\begin{eqnarray}
\chi' \propto \langle M^2 \rangle - \langle|M|\rangle^2
\label{47}
\end{eqnarray}
By taking the modulus of the magnetisation, one tries to minimise finite-size effects. Besides giving information about $T_c$, which is in agreement
with the results coming from the cumulants and the magnetisation, the susceptibility provides information on the critical exponent
$\gamma_s$. At $T_c$, one expects $\chi' \propto N^{\gamma_s/(2-\alpha)}$. We calculated $\chi'$ as a function of $N$ at a temperature
close to $T_c$. We find that $\gamma_s/(2-\alpha)$ is close to $0.47 \pm 0.03$, which is consistent with $\gamma_s=1$ using the earlier estimate of
$\alpha$.

\begin{figure}[htbp]
\centerline{\includegraphics[width=2.5in]{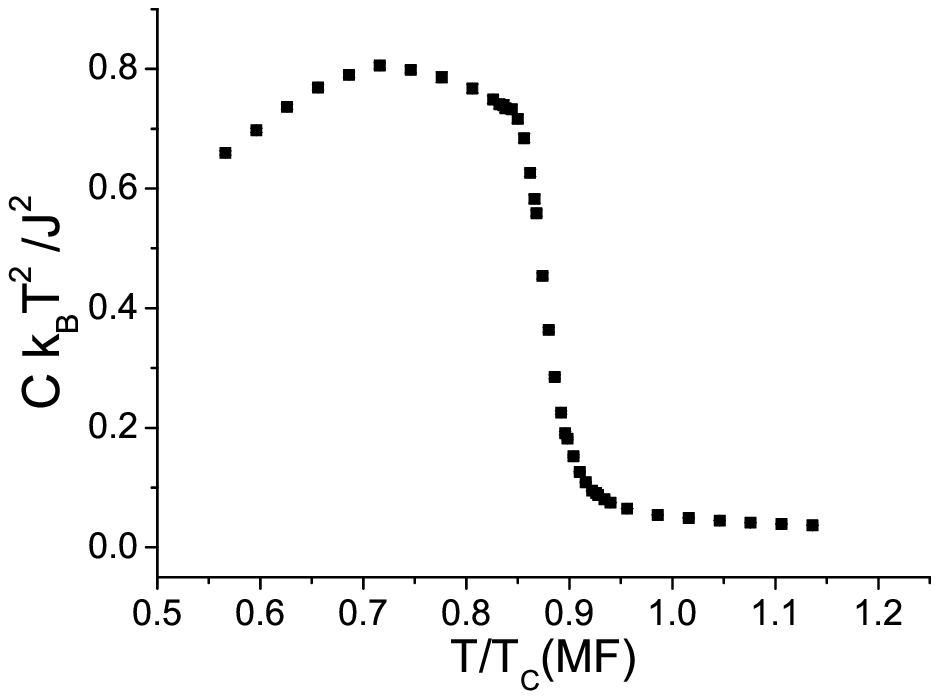}
\includegraphics[width=2.5in]{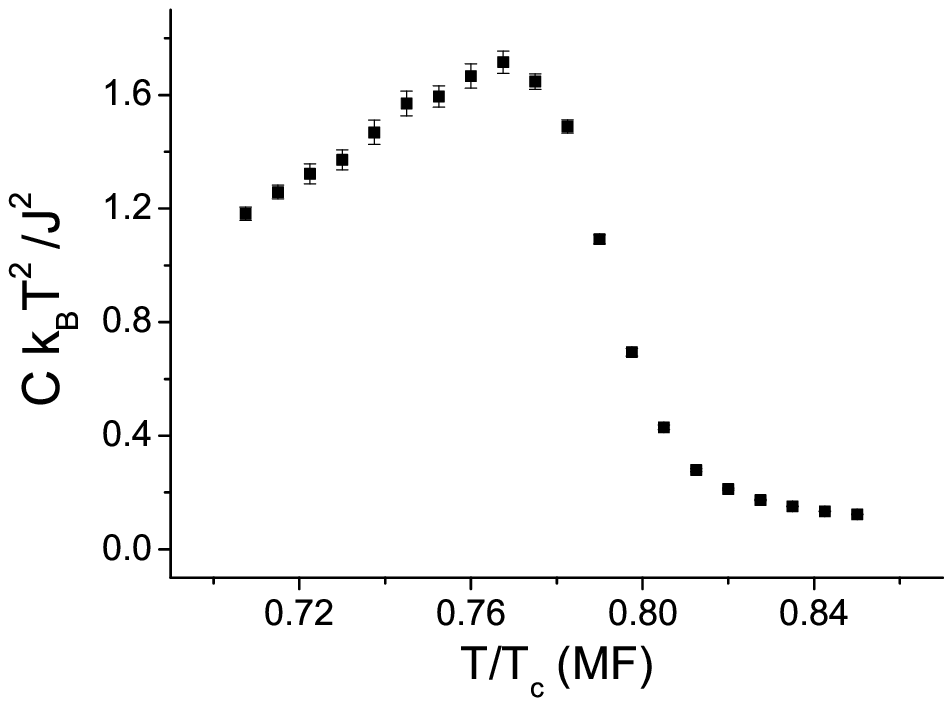}}
\caption{The specific heat as a function of temperature for a
network of 5600 nodes (SAN model , left) and for the modified
model with $\mu=1$ (right). Averages were taken over 400 samples.
The average connectivity is $Q=10$. } \label{fig9}
\end{figure}

The plots for the specific heat as a function of temperature (Fig. \ref{fig9}, left side) look similar to those obtained from the cavity method. The specific heat does not
diverge with size but saturates, which implies $\alpha \simeq 0$. However, the plots do not show the usual, mean field jump at $T_c$, but are in
agreement with the presence of logarithmic corrections. The maximum in the specific heat below $T_c$ that was found within the cavity
approach (Fig. \ref{fig3}) is also clearly visible in the data. When we change the value of $\mu$ from $1/2$ to $1$ we do find results consistent with the usual mean field behaviour, i.e. with
a jump at $T_c$ (see Fig. 10, right side and compare with Fig. 4). Indeed $\mu=1$ corresponds to $\gamma' \to \infty$, which corresponds to a network with a narrow degree distribution.

\begin{figure}[htbp]
\centerline{\includegraphics[width=2.5in]{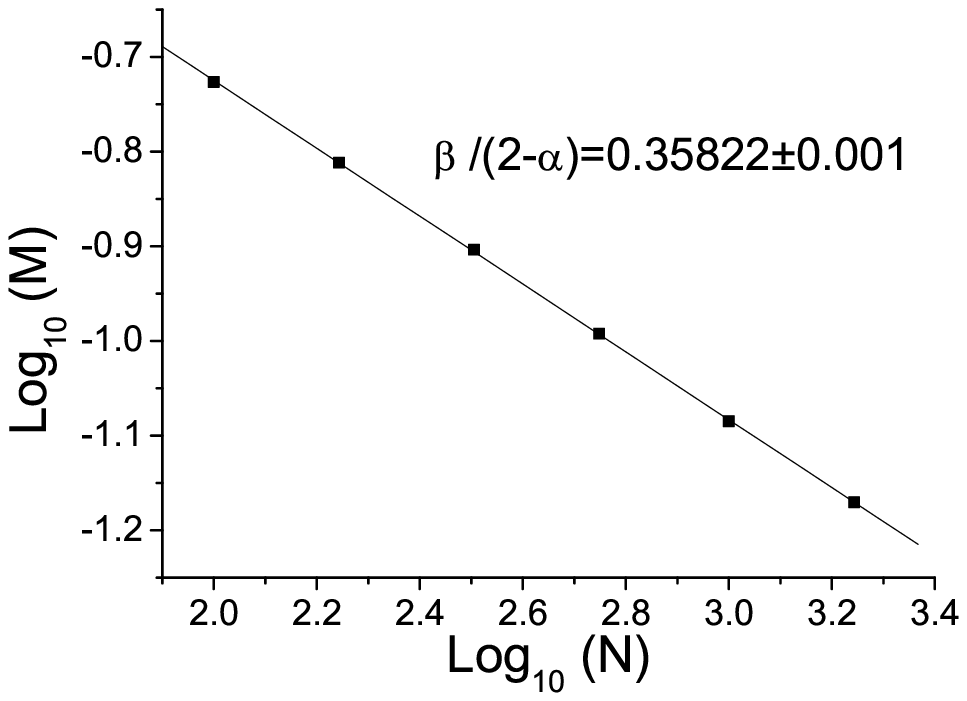}
\includegraphics[width=2.5in]{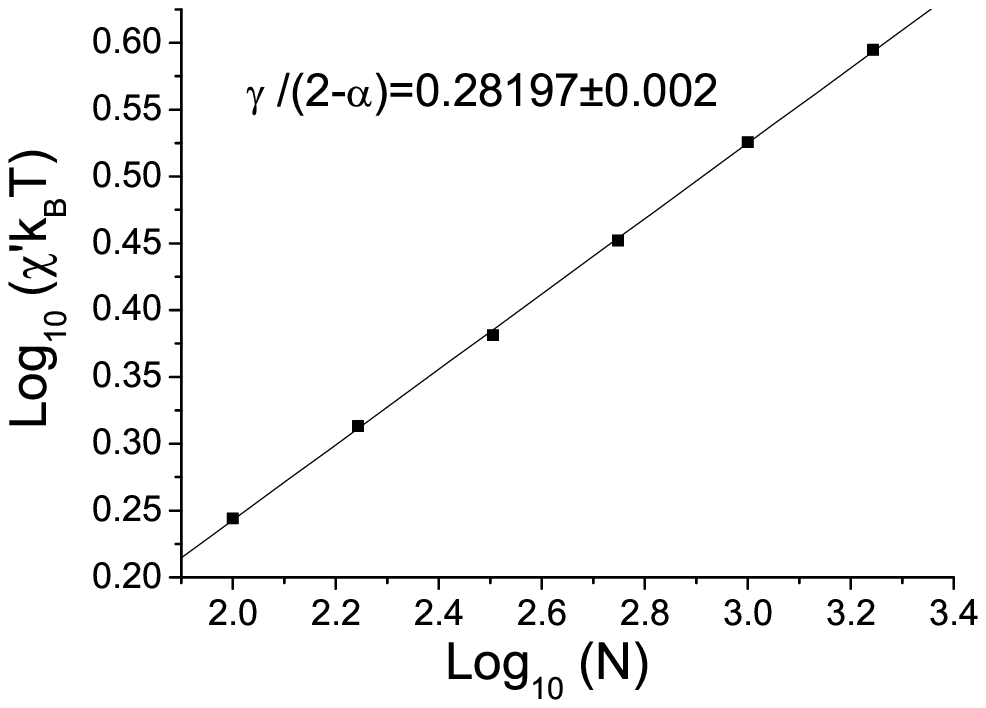}}
\caption{Log-log plot of the magnetisation (left) and the
susceptibility (right) versus $N$ at $T_c$ for the case $\mu=1/3,
Q=10$.} \label{fig10}
\end{figure}

\begin{figure}[htbp]
\centerline{\includegraphics[width=2.5in]{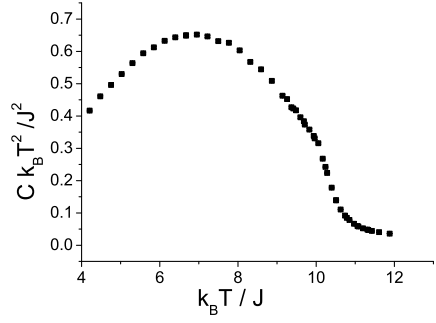}}
\caption{The specific heat as a function of temperature for a
network of 1000 nodes ($\mu=1/3$). The average connectivity is
$Q=10$. Notice the linear regime for $8 < k_B T/J < 10.2$ }
\label{fig11}
\end{figure}

We also performed a completely similar investigation of the network with $\mu=1/3$ and $Q=10$.
In Fig. \ref{fig10}, we show log-log plots of the magnetisation (left) and the susceptibility (right) as a function of $N$ at the numerically determined value of $T_c$. From these data we obtain
$\beta/(2-\alpha)=0.358\pm 0.001 $ and $\gamma_s/(2-\alpha)=0.282\pm 0.002$. These results are
consistent with the predictions coming from (\ref{12}): $\beta/(2-\alpha)=1/3$ and $\gamma_s/(2-\alpha)=1/3$. The remaining discrepancy is probably due to errors in the determination of $T_c$. We also obtained data for the specific heat, but they do not allow a precise determination of $\alpha$. From (\ref{12}), the prediction $\alpha=-1$ follows. This implies that the specific heat decreases linearly to its high-temperature background value.
The data shown in Fig. \ref{fig11} are consistent with this expectation.

\begin{figure}
\centerline{\includegraphics[width=2.2in]{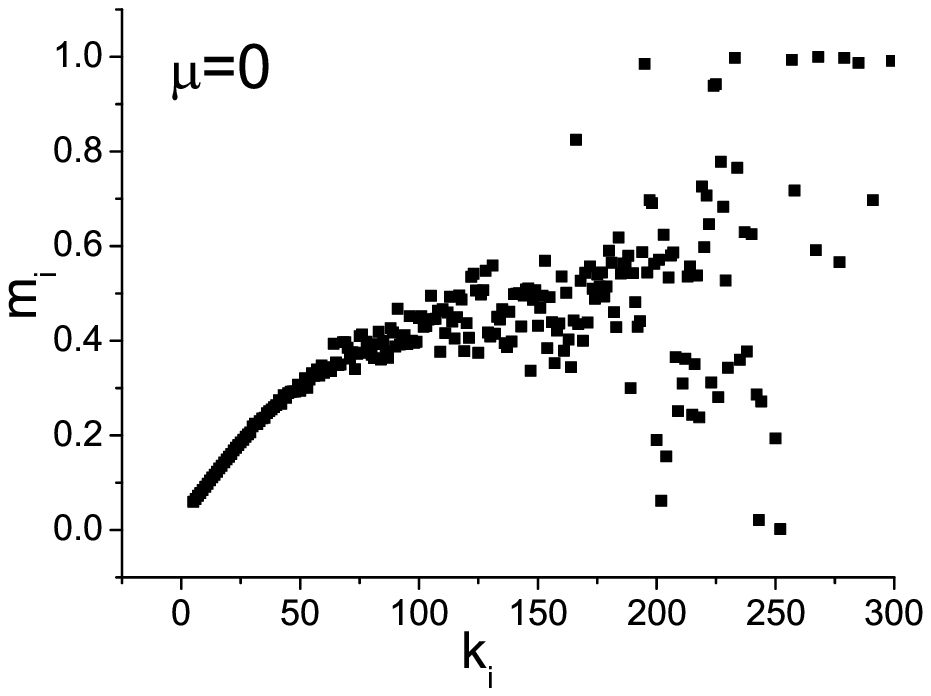}
\includegraphics[width=2in]{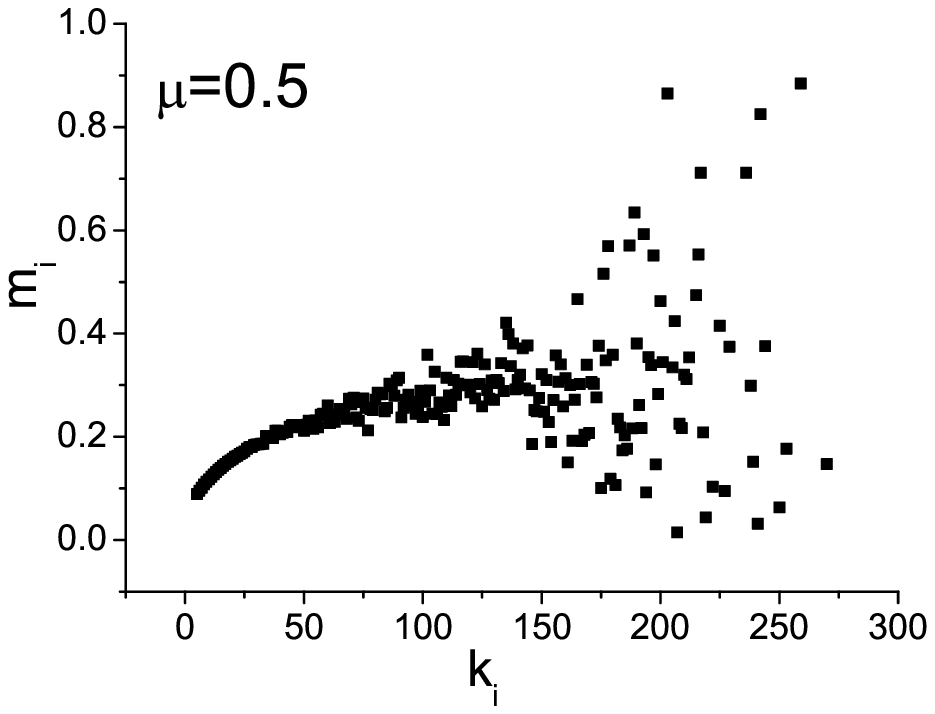}
\includegraphics[width=2in]{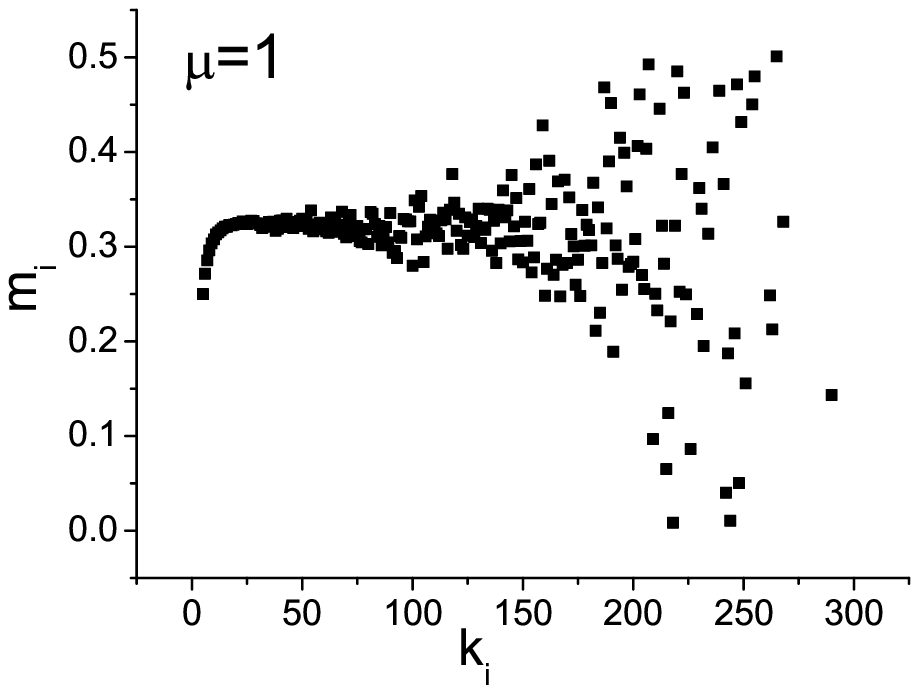}}
\caption{The magnetization of individual spins as a function of
degree for a network of 1000 nodes and different values of $\mu$.
The averages are made over 1000 networks. The values of $\mu$ and
the predicted $k$-dependence, respectively, is: $\mu=0$ - linear;
$\mu=0.5$ - square root; $\mu=1$ - constant. The temperature is
slightly below the respective critical temperatures.}
\label{fig12}
\end{figure}

From the simple mean field theory of section 2, it follows that the local order parameter $m_i \propto k_i^{1-\mu}$. We also used this relation in our derivation of
the BP approximation. In order to get further confirmation for this type of scaling, we checked it numerically.  Our results are shown in Fig. \ref{fig12}, for the cases
$\mu=0,\ 0.5$ and $1$ on a BA-network. For high values of the degree, the statistics is not very good,
but for smaller $k_i$-values the scaling (\ref{301}) seems to be well satisfied.

\begin{figure}[htbp]
\centerline{\includegraphics[width=2.5in]{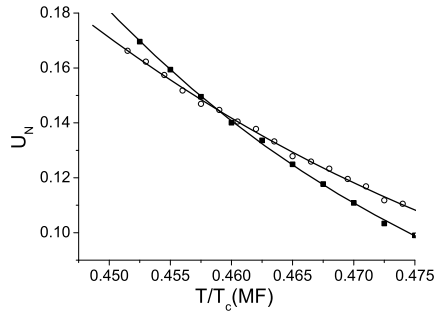}}
\caption{Binder cumulants as a function of temperature for a
network with $\gamma=3, Q=2$ and $\mu=1/2$. The figure shows a
clear intersection of the cumulants for $N=2000$ (circles) and
$N=4000$ (squares) indicating the presence of a critical point at
finite temperature. The averages were done over 6000 and 5000
realizations of the networks, respectively. The fit with a third
degree polynomial is also shown in the figure.} \label{fig13}
\end{figure}

We finally remark that for the SAN with $Q=2$, the Monte Carlo simulations indicate a zero critical temperature. Indeed, for this value of
$Q$, the BA network has the structure of a tree and therefore it cannot support order. This is in contrast with the results coming from
the replica approach. However, in that approach loops are not excluded and the network can be a
collection of
disconnected clusters, the largest of which determines the non-zero $T_c$ in
the large-$N$ limit. To test this, we generated a set of random (uncorrelated) networks with $Q=2$ and $\gamma=3$ by ascribing to each vertex a degree taken from the distribution $P(k)$. The links going out of the vertices are then randomly paired to create a network. The resulting network is indeed found to consist of disconnected parts.
In  Fig. \ref{fig13} we show Monte Carlo results for the Binder cumulant as a function of temperature for two  networks ($N=2000$ and $N=4000$) generated in this way. We took $\mu=1/2$. Clearly there is an intersection from which we obtain the finite-size estimate for the critical temperature. After extrapolation ($N \to \infty$) we obtain from these kind of data $k_BT_c/J= 0.92648$. For comparison, the BP prediction for $kT_c/J$ is $0.952$ and
the replica method
gives $0.946$ (Table 1).

\section{Nonequilibrium models}
In this section, we show that the basic result of this paper, (\ref{12}), is also true for a nonequilibrium model. In particular, we will investigate the contact process \cite{23} on a network. This process is a well known model originally introduced to describe the spreading of a disease in a population. Later, it was found to be related to directed percolation \cite{26} and became the prototype model used in the study of absorbing state phase transitions \cite{25}. Recently the contact process has also been applied in metapopulation ecology \cite{24}, but in this paper we will use the epidemiological language.

On a network we define the contact process as follows. Each vertex $i$ can be in two states that we denote as ill ($n_i=1$) and healthy ($n_i=0$). The dynamics of the model is given by a continuous time Markov process \cite{81}. An ill site can cure with a rate 1 (this fixes the time scale). A healthy site becomes ill with a rate that equals $\lambda$ times the number of ill neighbours, where on a network two vertices are called ´neighbours' if they are linked.

This model has an absorbing state: if all sites are healthy, they will stay healthy forever.
For an infinite system, there is
a phase transition at some $\lambda_c$. If $\lambda \leq \lambda_c$, the system will evolve to the absorbing state. For $\lambda > \lambda_c$, there will always be a finite density of ´ill' sites. This density is the order parameter of the model and it goes to zero if one approaches $\lambda_c$ from above.
Vespignani and coworkers \cite{19,20} showed that on a scale-free network
for $2 < \gamma \leq 3$, $\lambda_c=0$ and the exponents are $\gamma$-dependent.
For $3 < \gamma \leq 4$, $\lambda_c > 0$ and the critical exponents are again $\gamma$-dependent. Finally,
for $\gamma > 4$, $\lambda_c > 0$ and critical exponents assume the mean-field values for the contact process on a regular lattice. This scenario is reminiscent of that for the opinion formation model.

We next generalise the contact process by taking the infection rate as
\begin{eqnarray}
\lambda_{ij}= \lambda Q^{2\mu} (k_ik_j)^{-\mu}
\label{1001}
\end{eqnarray}
Using standard techniques from the theory of stochastic processes \cite{81}, one can show that $\rho_i=\langle n_i \rangle$ (where $\langle . \rangle$ in this paragraph denotes the average over histories of the stochastic process) obeys the exact equation
\begin{eqnarray}
\frac{d\rho_i}{dt} = - \rho_i + \sum_{l} \lambda_{il} \langle (1-n_i) n_l \rangle
\label{1002}
\end{eqnarray}
The sum runs over the neighbours of $i$. We now make two approximations similar to those performed in section 2. The first one, the
 {\it dynamical mean field approximation}, puts $\langle (1-n_i) n_l \rangle \approx \langle(1-n_i)\rangle \langle n_l \rangle = (1-\rho_i)\rho_l$.
Next, again following Bianconi \cite{7}, we replace the sum over the neighbours by a sum over all the nodes, and take for $\lambda_{ij}$ its average over all realisations of the network with a given set of degrees.
Then (\ref{1002}) becomes
\begin{eqnarray}
\frac{d\rho_i}{dt} = -\rho_i + (1-\rho_i) \sum_{j=1}^N [\lambda_{ij}] \rho_j
\label{1003}
\end{eqnarray}
Now (see section 2) $[\lambda_{ij}]= \lambda_{ij}p_{ij}$ with $p_{ij}=k_ik_j/(QN)$, the probability that nodes $i$ and $j$ are connected. This gives
\begin{eqnarray}
\frac{d\rho_i}{dt} = -\rho_i + (1-\rho_i) \frac{\lambda Q^{2\mu-1}}{N} k_i^{1-\mu} \sum_{j=1}^N k_j^{1-\mu} \rho_j
\label{1004}
\end{eqnarray}
We introduce $\Theta$
\begin{eqnarray}
\Theta = \frac{Q^{\mu-1}}{N} \sum_{j=1}^N k_j^{1-\mu} \rho_j
\label{1005}
\end{eqnarray}

From here on, we will assume that $\rho_i$ depends {\it only on the degree of $i$}. We can then write
for a large network
\begin{eqnarray}
\Theta= Q^{\mu-1} \sum_{k} P(k) k^{1-\mu} \tilde{\rho}_k
\label{1006}
\end{eqnarray}
where $\tilde{\rho}_k$ is the density of sites that are ill and have degree $k$.

Using (\ref{1005}) and this assumption, (\ref{1004}) becomes
\begin{eqnarray}
\frac{d\tilde{\rho}_k}{dt} = - \tilde{\rho}_k + \lambda (1-\tilde{\rho}_k) Q^u k^{1-\mu} \Theta
\label{1007}
\end{eqnarray}

We are interested in the {\it static properties}, i.e. the properties of the model in the long time limit. Then, from (\ref{1007}) we get
\begin{eqnarray}
\tilde{\rho}_k = \frac{\lambda Q^\mu k^{1-\mu} \Theta}{1 + \lambda Q^\mu k^{1-\mu} \Theta}
\label{1008}
\end{eqnarray}
Using (\ref{1006}), we find a self-consistency equation for $\Theta$
\begin{eqnarray}
\Theta = \lambda Q^{2\mu-1} \sum_k P(k) \frac{k^{2-2\mu} \Theta}{1+\lambda Q^\mu k^{1-\mu}\Theta}
\label{1009}
\end{eqnarray}

We can now in principle analyse this equation.

But as was the case for the opinion formation model, it is easier to transform away the $\mu$-dependence.
This is most easily done by rewriting (\ref{1009}) as an integral
\begin{eqnarray}
\Theta = A \lambda Q^{2\mu-1} \int_m^{\infty} k^{-\gamma}\frac{k^{2-2\mu} \Theta}{1+\lambda Q^\mu k^{1-\mu}\Theta} dk
\label{1010}
\end{eqnarray}
where $A$ is a normalisation constant and where  the power-law form for $P(k)$ is inserted.
If we change variables to $k'= Q^\mu k^{1-\mu}$, (\ref{10}) becomes (if $\mu \leq 1$)
\begin{eqnarray}
\Theta = A' \int_{m'}^{\infty} (k')^{\frac{1-\gamma}{1-\mu}} \frac{k'\Theta}{1+\lambda k' \Theta} dk'
\end{eqnarray}
where $A'$ is another constant and $m'=m^{1-\mu}Q^\mu$.

This is precisely the form (\ref{1010}) assumes for $\mu'=0$ provided we shift $\gamma$ to $\gamma'$
given by
\begin{eqnarray}
\gamma'=\frac{\gamma-\mu}{1-\mu}
\end{eqnarray}
Thus, the same relation holds as for the static Ising case \cite{27}. But notice that now, taking $\mu=1/2$ and a Barabasi-Albert network (i.e. the epidemiological version of the SAN), leading to
$\gamma'=5$, we are in a situation where mean-field theory should hold {\it without logarithmic corrections}.

The relation  (\ref{12}) thus appears to be quite general and one may wonder whether there are any exceptions to it.
We first discuss a recently proposed realization of the
Bak-Tang-Wiesenfeld sandpile model on scale-free networks
\cite{22}. In this model an avalanche can be generated through
the following dynamical rules: 1) at each time step, a grain is
added at a randomly chosen network node $i$; 2) if the height
$h_i$ at node $i$ exceeds a given threshold $z_i$, it becomes
unstable and an integer number of grains, $n[z_i]$, at the node
topple, with $n[z_i]-1 < z_i \leq n[z_i]$, to randomly chosen
$n[z_i]$ nodes among $k_i$ adjacent ones; 3) whenever adjacent
nodes become unstable toppling takes place also there, on all
unstable nodes in parallel, and the avalanche continues until
there are no unstable nodes left. In the proposed version of the
model \cite{22}, a parameter $\eta$ is introduced and the
threshold $z_i$ of node $i$ is taken to be
\begin{equation}
z_i = k_i^{1-\eta}
\end{equation}
Note that for $\eta =0$ the model features a simple
degree-dependent threshold, and in the extended model $k$ is
replaced by $k^{1-\eta}$. Note that this is reminiscent of the
transformation $ k \rightarrow k^{1-\mu}$ which we discussed in
other contexts in this paper.

A key quantity in the description of the avalanche dynamics by
mapping each avalanche to a tree, is the branching probability
$P_b(n)$ that a node, which receives a grain from a neighbour,
generates $n$ branches. This probability consists of two factors
\cite{22}. If branching occurs for a given node $i$, $n[z_i]$
branches are generated. Therefore, the first factor, $P_1(n)$, is
the probability that $n$ coincides with $n[z_i]$ for a node $i$
already connected to the tree. The second factor is the
probability that the height $h_i$ takes the value $n -1$ at the
moment that node $i$ receives a grain from a neighbour. This
probability, $P_2(n)$, is not important for us here, but in the
case of independent random heights ${0,1, ..., n-1}$ in the
inactive state of the sandpile, it equals $1/n$. If we would adopt
a continuum approximation, treating $n$ as a real number, and
approximate $n[z_i]$ by $z_i$, we would arrive at the result
\begin{equation}
P_b(n) = (n^{1/(1-\eta)} P(n^{1/(1-\eta)})/Q)P_2(n),
\end{equation}
where the first factor, $P_1(n)= k P(k)/Q$, is the probability
that the first neighbour of a node has degree $k$, with in our
case $k = n^{1/(1-\eta)}$. This probability differs from $P(k)$ in
that it presupposes that the generating node is already connected
to the tree.

If we compare this $P_b(n)$ with its counterpart for the model
with $\eta = 0$, we easily observe that the extended model is
equivalent to the basic model with $\eta' = 0$, provided the
topological exponent is transformed according to
\begin{equation}
\gamma' = \frac{\gamma - \eta}{1-\eta}
\end{equation}
This is in agreement with our general exponent relation, but, as
we shall show now, and as Goh {\em et al.} already found \cite{22}
this result is incorrect.

The correct exponent relation is found when taking into account
properly that $n$ is always an integer, and that all thresholds in
the interval $n-1 < z_i \leq n$ contribute to the probability of
generating $n$ branches. $P_1(n)$ thus consists of a sum, or, for
simplicity and without loss of relevant precision, an integral, so
that
\begin{equation}
P_b(n) = (\int_{(n-1)^{1/(1-\eta)}}^{n^{1/(1-\eta)}}kP(k)dk
/Q)P_2(n),
\end{equation}
from which follows the correct relation
\begin{equation}
\gamma' = \frac{\gamma - 2 \eta}{1-\eta},
\end{equation}
which was already obtained by Goh {\em et al} \cite{22}. We
conclude that an exception to our general exponent relation arises
here due to the discrete (integer) character of the toppling
process. Indeed, the general relation is recovered in a (rough)
continuum approximation of the problem.

A second exception to our general exponent relation is provided by
the study of Dezs\"o and Barab\'asi \cite{22bis} of disease
spreading on a scale-free network. The model they consider starts
from the usual contact process, for which it is known that the
epidemic threshold $\lambda_c$ vanishes for $\gamma \leq 3$. This
is similar to the divergence of the critical temperature for an
Ising model with constant interactions on a scale-free network. We
have seen that topology-dependent interactions are a way to get
around this and we have discussed this in detail also for the
contact process.

However, Dezs\"o and Barab\'asi provide an alternative means of
rendering the epidemic threshold finite and thus offering new
avenues for controlling diseases (e.g., eradicating viruses).
Their strategy consists of curing the hubs with a probability that
scales with the degree $k$ of a node as $k^\alpha$. Note that
$\alpha=0$ corresponds to the usual model in which all nodes are
cured with the same probability.

As a result Eq. (\ref{1010}) takes the modified form
\cite{22bis}
\begin{equation}
\Theta = A \lambda Q^{-1}\int_m^{\infty} k^{-\gamma} \frac{k^2
\Theta}{k^\alpha + \lambda k \Theta} dk,
\end{equation}
from which we can derive the following exponent relation
\begin{equation}
\gamma' = \frac{\gamma - 2 \alpha}{1-\alpha},
\end{equation}
with $\gamma'$ the effective topological exponent for an
equivalent model with degree-independent curing rate ($\alpha' =
0$). Curiously, this new relation is akin to that of the previous
exception (sandpile model), but this coincidence has to our
insight no significance.

\section{Discussion}
\noindent In this paper, we investigated the critical properties
of an Ising model and a contact process with topology-dependent
interactions on scale-free networks. The interaction strength
between two spins, or the infection rate between two individuals,
was assumed to be proportional to $(k_i k_j)^{-\mu}$. We have
developed mean-field theories for these models from which our main
result follows: the critical behaviour can  always be related to
that of the corresponding model with homogeneous couplings
($\mu'=0$) on a network with a modified degree distribution
$\gamma'$, where $\gamma'$ is given by equation (\ref{12}). Due to
the small-world property of scale-free networks, mean field theory
is generally believed to be exact. This expectation is found to be
true also in the present case. We performed extensive numerical
calculations for the Ising model on a BA-network, focusing on the
cases $\mu=1/2$ (the SAN, which was the original motivation for
the present work) and $\mu=1/3$. We used two techniques:
´numerically exact' studies with the cavity approach, and Monte
Carlo simulations. The first approach assumes the absence of loops
in the network, which is a good approximation for a BA-network.
The Monte Carlo calculations approximate the true thermal average
by an average over a finite set of well chosen spin
configurations. Within their numerical accuracy, both approaches
give the same results. More importantly, they provide strong
evidence that the exponent relation (\ref{12}) is correct.

For static network ensembles it is possible to obtain an exact equation for the location of the critical temperature from the replica approach, see equations (\ref{34}) and (\ref{35}). A simpler equation for this quantity, (\ref{305}), can be derived from the Bethe-Peierls approximation. From the numerical values listed in table 1, it can be seen that this simple aproach gives results that are accurate to better than one percent. From the table one also observes that the differences in the critical temperature as obtained from the different approaches are very small for $Q \geq 4$.

Besides the contact process, the behaviour of several other (non) equilibrium models have been studied on scale-free networks. In particular, we mention here diffusion-annihilation \cite{21} and the Bak-Sneppen model \cite{255}. In both cases, the critical behaviour shows a $\gamma$-dependence that is reminiscent of that of the Ising model and the contact process. It remains to be investigated whether appropriately extended versions of these models obey the relation (\ref{12}) or whether they follow modified transformation laws as we found to be the case for the sandpile model of ref. 43 or the modified contact process of ref. 44.

\section*{Acknowledgements} \noindent We would like to thank T.
Coolen, F. Igl\'{o}i, M. M\"{u}ller, M. Baiesi and D. Stauffer for
useful discussions. J.O. Indekeu thanks the FWO-Vlaanderen for
financial support through the project G.0222.02. M. Leone would
like to acknowledge the EVERGROW EU Integrated Project for
research funding and the Katholieke Universiteit Leuven for travel
funding and for kind hospitality. I. P\'{e}rez Castillo
acknowledges support from EPSRC under grant GR/R83712/01.


\begin{thebibliography}{000}
\bibitem{-2}
R. Albert and A.-L. Barab\'{a}si, {\it Rev. Mod. Phys.} {\bf 74}, 47 (2002).
\bibitem{-1}
S.N. Dorogovtsev and J.F.F. Mendes, {\it Evolution of networks: from biological nets to the internet and WWW}, Oxford University Press (2003).

\bibitem{-5} B. Bollob\'{a}s, {\it Random Graphs}, Cambridge University Press (2001).
\bibitem{-4} D.J. Watts and S.H. Strogatz (1998), {\it Nature} {\bf 393} 440 (1998).
\bibitem{-3} S. Maslov, K. Sneppen and U. Alon in {\it Handbook of graphs and networks} edited by S. Bornholdt and H.G. Schuster, Wiley (2003), p. 168.
\bibitem{101}
R. Pastor-Satorras and A. Vespignani, {\it Evolution and structure of the internet: a statistical physics approach}, Cambridge University Press (2004).
\bibitem{102} H. Jeong, B. Tombor, R. Albert, Z.N. Oltvai and A.-L. Barab\'{a}si, {\it Nature} {\bf 407}, 651 (2000).
\bibitem{104} R.V. Sol\'{e} and A. Munteanu, {\it Europhys. Lett.} {\bf 68}, 170 (2004).
\bibitem{103}
L.A.N. Amaral and J.M. Ottino, {\it Eur. Phys. J. B} {\bf 38}, 147 (2004).
\bibitem{1}
A.-L. Barab\'{a}si and R. Albert, {\it Science} {\bf 286}, 509 (1999).
\bibitem{105}
S. Galam, Y. Gefen, and Y. Shapir, J. Math. Sociology {\bf 9}, 1 (1982); K. Sznajd-Weron and J. Sznajd, Int. J. Mod. Phys. C {\bf 11}, 1157 (2000).
\bibitem{0}
A. Aleksiejuk, J.A. Holyst, and D. Stauffer, {\it Physica A} {\bf 310}, 260 (2002).
\bibitem{2}
S.N. Dorogovtsev, A.V. Goltsev, and J.F.F. Mendes, {\it Phys. Rev. E} {\bf 66}, 016104 (2002).
\bibitem{3}
M. Leone, A. V\'{a}zquez, A. Vespignani, and R. Zecchina, {\it Eur. Phys. J. B} {\bf 28}, 191 (2002).
\bibitem{4}
A.V. Goltsev, S.N. Dorogovtsev, and J.F.F. Mendes, {\it Phys. Rev. E} {\bf 67}, 026123 (2003).
\bibitem{5}
F. Igl\'{o}i and L. Turban, {\it Phys. Rev. E} {\bf 66}, 036140 (2002).
\bibitem{6}
J.O. Indekeu, {\it Physica A} {\bf 333}, 461 (2004).
\bibitem{0p}
C.V. Giuraniuc, J.P.L. Hatchett, J.O. Indekeu, M. Leone, I. P\'{e}rez Castillo, B. Van Schaeybroeck and C. Vanderzande, {\it Phys. Rev. Lett.} {\bf 95}, 098701 (2005).
\bibitem{7}
G. Bianconi, {\it Phys. Lett. A} {\bf 303}, 166 (2002).
\bibitem{11}
J.S. Yedidia in {\it Advanced Mean field methods: theory and practice}, edited by M. Opper and D. Saad, MIT press (2001).
\bibitem{8}
M. M\'{e}zard, G. Parisi and M.A. Virasoro, {\it Spin glass theory and beyond}, World Scientific (1987).
\bibitem{9}
M. M\'{e}zard and G. Parisi, {\it Eur. Phys. J. B} {\bf 20}, 217 (2001)
\bibitem{10}
M. M\'{e}zard and G. Parisi, {\it J. Stat. Phys.} {\bf 111}, 1 (2003).
\bibitem{12}
M. Newman, {\it Phys. Rev. Lett.} {\bf 89}, 208701 (2002).
\bibitem{13}
A. V\'{a}zquez and M. Weight, {\it Phys. Rev. E} {\bf 67}, 027101 (2003).
\bibitem{14}
G. Bianconi and A. Capocci, {\it Phys. Rev. Lett.} {\bf 90}, 078701 (2003).
\bibitem{15}
H.D. Rozenfeld, J.E. Kirk, E.M. Bollt and D. ben-Avraham, {\it J. Phys. A} {\bf 38}, 4589 (2005).
\bibitem{777}
R.J. Baxter, {\it Exactly solved models in statistical mechanics}, Academic Press (1982).
\bibitem{16}
R. Kikuchi, {\it Phys. Rev.} {\bf 81}, 988 (1951).
\bibitem{17}
K. Y. Wong  and D. Sherrington, {\it J. Phys. A} {\bf 21}, L459 (1988)
\bibitem{201}
K. Binder, {\it Z. Phys. B} {\bf 43}, 119 (1981).
\bibitem{202}
K. Binder, {\it Phys. Rev. Lett.} {\bf 47}, 693 (1981).
\bibitem{203}
A.M. Ferrenberg and D.P.Landau, {\it Phys. Rev. B} {\bf 43}, 5081 (1991).
\bibitem{18}
K. Binder in {\it The Monte Carlo method in the Physical Sciences} edited by J.E. Gubernatis {\it AIP Conference Proceedings}, Vol 690, Springer Verlag (2003).
\bibitem{23}
T. E. Harris, {\it Ann. Prob.} {\bf 2}, 969 (1974).
\bibitem{26}
P. Grassberger and A. de la Torre, {\it Ann. Phys.} (NY) {\bf 122}, 373 (1979).
\bibitem{25}
R. Dickman  in {\it Nonequilibrium statistical mechanics in
one dimension }edited by V. Privman , Cambridge University Press (1997).
\bibitem{24}
V. Vuorinen, M. Peltom\"{a}ki, M. Rost and M.J. Alava, {\it Eur. Phys. J. B} {\bf 38}, 261 (2004).
\bibitem{81}
N.G. Van Kampen, {\it Stochastic processes in physics and chemistry}, North-Holland (1992).
\bibitem{19}
R. Pastor-Satorras and A. Vespignani, {\it Phys. Rev. E} {\bf 63}, 066117 (2001).
\bibitem{20}
R. Pastor-Satorras and A. Vespignani in {\it Handbook of graphs and networks} edited by S. Bornholdt and H.G. Schuster, Wiley (2003).
\bibitem{27}
This extension of the exponent relation to nonequilibrium models
was independently obtained in M. Karsai, R. Juh\'{a}sz and F.
Igl\'{o}i, {\it Phys. Rev. E} {\bf 73}, 036116 (2006)
\bibitem{22}
K.-I. Goh, D.-S. Lee, B. Kahng and D. Kim, {\it Physica A} {\bf
346}, {93} (2005).
\bibitem{22bis}
Z. Desz\"{o} and A.-L. Barab\'{a}si, {\it Phys. Rev. E} {\bf 65},
055103(R) (2002).
\bibitem{21}
M. Catanzaro, M. Bogu\~{n}a and R. Pastor-Satorras, {\it Phys. Rev. E} {\bf 71}, 056104 (2005).
\bibitem{255}
S. Lee and Y. Kim, {\it Phys. Rev. E} {\bf 71}, 057102 (2005).
\end{thebibliography}
\end{document}